\documentclass[12pt]{article}
\usepackage{authblk}
\usepackage[comma,numbers,square,sort&compress]{natbib}
\usepackage{graphicx}
\usepackage[margin=1in]{geometry}
\usepackage[colorlinks,linkcolor=blue,citecolor=blue,urlcolor=blue]{hyperref}
\usepackage{xspace,setspace}
\usepackage{lineno}
\newcommand{\efig}{Extended Data Fig.}
\newcommand{\smv}{Supplementary Video}
\newcommand{\kms}{km~s$^{-1}$}

\newcommand{\aap}{    {\it Astron. Astrophys.}}

\newcommand{\aapr}{   {\it Astron. Astrophys. Rev.}}
\newcommand{\ag}{     {\it Ann. Geophys.}}

\newcommand{\apj}{    {\it Astrophys. J.}}
\newcommand{\apjl}{   {\it Astrophys. J. Lett.}}

\newcommand{\araa}{   {\it Annu. Rev. Astron. Astrophys.}}

\newcommand{\jgr}{    {\it J. Geophys. Res.}}
\newcommand{\mnras}{  {\it Mon. Not. R. Astron. Soc.}}

\newcommand{\solphys}{{\it Sol. Phys.}}
\newcommand{\ssr}{    {\it Space Sci. Rev.}}
\newcommand{\prl}{    {\it Phys. Rev. Lett.}}
\newcommand{\lrsp}{   {\it Living Rev. Sol. Phys.}}
\newcommand{\ija}{    {\it Int. J. Astrobiol.}}
\newcommand{\nastro}{ {\it Nat. Astron.}}
\newcommand{\ncomms}{ {\it Nat. Commun.}}

\newcommand{\sciadv}{ {\it Sci. Adv.}}
\newcommand{\raa}{    {\it Res. Astron. Astrophys.}}

\begin{document}

\title{\Large \bfseries Multi-Viewpoint Observation of a Failed Prominence Eruption on the Sun}

\author[1,*]{Tingyu Gou}
\author[1]{Katharine K. Reeves}
\author[2,3]{Peter R. Young}
\author[4]{Astrid M. Veronig}
\author[5]{Xingyao Chen}
\author[5]{Sijie Yu}
\author[5]{Bin Chen}
\author[6]{Bin Zhuang}

\affil[1]{Center for Astrophysics $|$ Harvard $\&$ Smithsonian, Cambridge, MA, USA}
\affil[2]{NASA Goddard Space Flight Center, Greenbelt, MD, USA}
\affil[3]{Department of Mathematics, Physics and Electrical Engineering, Northumbria University, Newcastle upon Tyne, UK}
\affil[4]{Institute of Physics \& Kanzelh\"{o}he Observatory for Solar and Environmental Research, University of Graz, Graz, Austria}
\affil[5]{Center for Solar-Terrestrial Research, New Jersey Institute of Technology, Newark, NJ, USA}
\affil[6]{Institute for the Study of Earth, Oceans, and Space, University of New Hampshire, Durham, NH, USA}
\affil[*]{E-mail: \url{tingyu.gou@cfa.harvard.edu}}

\date{March 25, 2026}
\maketitle

\begin{abstract}
Solar eruptions are sudden ejections of coronal mass and magnetic fields accompanied by intense energy release. The eruptive structure does not always erupt successfully, but sometimes fails to escape the Sun after initiation. The failure of an eruption, however, provides an invaluable opportunity for understanding the intricate mechanism of eruptions. We present a comprehensive observation of a failed prominence eruption on the Sun, taking advantage of multi-viewpoint and multi-messenger imaging. Simultaneous off-limb and on-disk observation gives evidence of magnetic reconnection processes occurring at different sites during the flare. Particularly, in addition to the standard flare reconnection behind the eruption, strong external reconnection occurs on the erupting flux rope, evidenced by a wealth of signatures via multi-wavelength imaging and spectroscopy. The two reconnection processes may play contrasting roles in the flux rope's acceleration and compete in altering the magnetic flux in the rope. As the high rate of external reconnection proceeds, the flux rope and embedded prominence decelerate noticeably and fail to erupt into the heliosphere, under strong magnetic confinement of overlying fields. Our results illustrate a well-defined physical picture for solar eruptive activities and provide insight into the lack of coronal mass ejections found in other solar-type stars.
\end{abstract}

The Sun is the closest and only star for which we can image the details of dynamic activities. The solar atmosphere, filled with magnetized plasma, provides natural experiments for diagnosing the magnetic processes that govern throughout the universe. Among them, magnetic eruptions are one of the most spectacular phenomena, often accompanied by prominence eruptions and energetic flares \cite{Priest2002}. Eruptions on the Sun can throw out massive coronal plasmas along with configured magnetic fields, named coronal mass ejections (CMEs) \cite{Chen2011,Webb2012}, into interplanetary space, causing extreme space weather effects. Eruptions are also expected to occur on other stars, which can have hazardous effects on exoplanets orbiting the host star such as an alteration of the planet's magnetosphere and outer radiation belt or even the removal of their atmospheres \cite{Airapetian2020,lugaz2016}. However, signatures of stellar CMEs are rare, in contrast to the large number of stellar flares that have been recorded \cite{Davenport2016,Vida2019,Veronig2021}. In addition to the detection feasibility, the rare detection of stellar CMEs may imply a high rate of failed or confined eruptions on late-type stars \cite{Odert2017}. Studies on the mechanism of solar eruptions provide valuable insights into the universal magnetic process.

Theoretical models have revealed various mechanisms for solar eruptions, including ideal magnetohydrodynamic (MHD) instabilities and resistive magnetic reconnection, which act as either a trigger or a driver in the eruptive process \cite{Green2018}. The eruption onset can be initiated, for example, by magnetic reconnection inside a sheared core field or at the coronal null point in a multiflux system \cite{Moore2001,Antiochos1999}, or by the ideal kink instability of a magnetic flux rope when its field lines show a large degree of twist \cite{Toeroek2004}. The magnetic flux rope consisting of helical magnetic field lines has been widely accepted as a fundamental structure in solar eruptions, and its dynamic evolution is key for the eruption process \cite{Gou2019,Liu2020}. Upon the initiation of the flux rope eruption, continuous acceleration drives the eruption, which can be usually facilitated by magnetic reconnection occurring at the current sheet (CS) in the wake of the rope \cite{Lin2000,Gou2020}, or by the torus instability when the rope experiences a sufficiently fast decay of overlying magnetic fields at a large coronal height \cite{Bateman1978,Kliem2006}. In observation of solar cases, these different physical processes often occur simultaneously and are coupled closely, which has consistently posed a difficulty in understanding the relevance of the different contributions. However, the eruptive structure may also experience a failed eruption process, during which the prominence or magnetic flux rope is activated to rise, but the eruption process is halted in the low corona without material or magnetic structure escaping the Sun. Without an immediate escape, the failure of an eruption offers an alternative and favorable opportunity to investigate the physical mechanisms of solar and stellar eruptions, i.e., how different physical processes govern the genesis of flares and escape of CMEs.

Studies have suggested several notable factors that affect the CME productivity during solar flares, including the degree of non-potentiality in the core field (e.g., sufficient magnetic free energy) and the constraining effect of background field \cite{Falconer2002,Wang2007,Sun2015}. Numerical experiments have found that the strength and decay of background field significantly affect the eruption \cite{Torok2005,Fan2007}, and thus strong overlying fields can lead to failed eruptions. Observations of failed eruption cases have revealed the influence of complex physical conditions and processes, such as kink motion, asymmetric confinement, interaction with surrounding loops, and complex magnetic configuration \citep{Ji2003,Liu2009,Chen2023,Karpen2024}. However, there is still a lack of comprehensive understanding of detailed dynamics and underlying mechanisms for failed eruptions, primarily due to limited observations capable of capturing both the eruptive dynamics and the coronal magnetic field at the same time. The launch of multiple solar missions and their coordinated observations have made such investigations possible.

We present observations of a failed prominence eruption accompanied by an intense flare on the Sun. The event has stereoscopic imaging from different viewpoints, including the Solar Dynamics Observatory (SDO) in the Sun-Earth line and Solar Orbiter (SolO; Fig.~\ref{fig:f1_overview}; Methods), which provides both off-limb and on-disk observations featuring the coronal dynamics and low-atmosphere characteristics simultaneously. In addition, the event involves coordinated multi-wavelength observations from various instruments on board satellites from the Sun-Earth line and from ground-based observatories, including UV/EUV imaging from the Atmospheric Imaging Assembly (AIA) on board SDO, X-ray imaging from the X-Ray Telescope (XRT) on Hinode, UV/EUV spectroscopy from the Interface Region Imaging Spectrograph (IRIS) and the EUV Imaging Spectrometer (EIS) on Hinode, and microwave imaging spectroscopy from the Expanded Owens Valley Solar Array (EOVSA) on the ground (Methods). The multi-viewpoint and multi-messenger observation reveals a wealth of details of physical processes occurring during the flare.

\section*{Results}

The event occurred at the west solar limb from Earth's perspective on 2024 March 30 (Fig.~\ref{fig:f1_overview}, \efig~1). It produced an intense solar flare with a high peak in the soft X-ray (SXR) flux at 21:16~UT, classified as an M9.4 class. The prominence that contains dense, cool material rises at about 21~UT (see \smv~1). Hot, tenuous plasmas emitting EUV and SXRs rise impulsively together with the prominence (\efig~1), serving as an indication of the magnetic flux rope structure that supports the prominence mass \citep{gibson2006,Zhang2012}. However, the impulsive rise is suspended soon after with a lot of prominence material falling back, and the hot flux rope structure becomes faint in the low solar corona following the deceleration (\efig~2; \smv~1). No large-scale CME structure is observed propagating into interplanetary space after the flare, except for a diffusive front that dissipates quickly in the outer corona (\smv~2; Methods). Various observations by multiple satellites verify the confinedness of this flare without ambiguity (Methods).

The flare is observed on the solar disk from SolO's viewpoint, which is separated from Earth by 44$^\circ$ in longitude (Fig.~\ref{fig:f1_overview}). The active region (AR) exhibits a large longitudinal extent of about 30$^\circ$, and almost half of the photospheric counterpart is occulted behind the SDO/AIA limb when the event occurs (\efig~3). The photospheric magnetogram prior to the flare shows a complex multipolar configuration, with four main magnetic polarities distributed in different longitudes (Fig.~\ref{fig:f2_AR}; referred to as P1/N1 in the AR center and P2/N2 at the two sides). A potential magnetic field extrapolation gives evidence of a magnetic X-point with a guide field in the corona, where magnetic separatrices intersect, producing a favorable site for magnetic reconnection (Methods; \efig~4). The X-type null point is located at $\sim$46~Mm above the photosphere, above the magnetic polarity inversion line (PIL) separating P1 and N1 in the AR center, embedded below the large-scale overlying magnetic field lines connecting P2 and N2 on the periphery. In EUV images, the prominence is observed to rise up from the core of the AR in the initial phase, right below the magnetic X-point (Fig.~\ref{fig:f2_AR}, \efig~3; Methods). The overall magnetic topology resembles the configuration of the magnetic breakout model in a multipolar magnetic system \cite{Antiochos1999,Lynch2008}. As such, during the eruption, magnetic reconnection not only occurs at the vertical CS beneath the rising flux rope, referred to as flare reconnection as described in the standard flare model \cite{Lin2000,Shibata2011}, but also occurs at the CS at the coronal null above the flux rope \cite{Karpen2012}. The post-reconnection configuration is demonstrated in EUV observations after the flare impulsive phase, where chromospheric footpoint brightenings are formed in all four main magnetic polarities, including in the remote magnetic polarities P2 and N2 that are initially inactive (Fig.~\ref{fig:f2_AR}, \efig~3). 

In the observations, we have found several pieces of evidence for magnetic reconnection occurring during the flare, including both below and above the flux rope. As the prominence and flux rope continuously rise, magnetic reconnection occurs at the flare CS behind the eruption, producing high-temperature, growing post-flare loops at low altitudes near the limb, as seen from AIA and XRT (\efig~1, \smv~1). The Spectrometer/Telescope for Imaging X-rays (STIX) onboard SolO records flare HXR emissions from accelerated electrons with energies up to 100 keV, showing non-thermal, high-energy HXR sources from the footpoints of post-flare loops in the AR center and thermal emissions from the flare loop top (Fig.~\ref{fig:f3_rec}b,d; Methods). Microwave emission observed by EOVSA on Earth shows a similar temporal evolution (e.g., at 10 GHz) as the non-thermal HXR emissions, indicating that there are accelerated particles around the flare CS and post-flare loops (Fig.~\ref{fig:f3_rec}c,d, \efig~5; Methods).

Besides the well-known flare reconnection process below the flux rope, we have also found clear signatures for the external reconnection occurring above the flux rope. For example, at $\sim$21:13~UT, a strong external reconnection process is evident in multiple observations (Fig.~\ref{fig:f3_rec}): a cusp-shaped structure suddenly appears at a high altitude in XRT and AIA hot channels, containing high-temperature plasmas $>$10~MK, indicative of a newly-formed hot reconnection outflow; the footpoint on the visible disk of AIA, which is located in the remote positive magnetic polarity P2 and is distinctly apart from the eruption in the AR center, exhibits intense brightening; some prominence material starts to detach from the main body of the erupting flux rope, suggestive of a transfer of magnetic flux in the flux rope resulting from the external reconnection. Moreover, EOVSA observes two short-lived bright radio bursts at frequencies of $\sim$2--6 GHz, which are clearly displaced from the gyrosynchrotron continuum emission associated with the flare reconnection, indicating coherent microwave emission from accelerated particles around the flux rope and the detached prominence (Fig.~\ref{fig:f3_rec}c, \efig~5; Methods). The two intense bursts show rapid drifts to higher frequencies in the spectrum (\efig~5c), indicative of accelerated electrons moving into a region with a higher density or a stronger magnetic field. Comparing the imaging observation and magnetic configuration of the AR, we find that at this time, the front of the rising prominence and flux rope approaches the height of the magnetic null point in the corona (Fig.~\ref{fig:f3_rec}; Methods). The close timing and spatial correlations suggest that strong external reconnection occurs above the erupting magnetic flux rope. This reconnection probably involves the magnetic field line of the twisted flux rope itself, which contains a strong magnetic field strength and results in multiple noteworthy features in the observation. This reconnection is also associated with a sudden drop in the rising velocity of both the prominence and flux rope (Fig.~\ref{fig:f3_rec}d). From then on, the impulsive eruption starts to slow down (see the kinematic evolution in \efig~2).

Another noteworthy feature for the external magnetic reconnection is a pair of cusp-shaped outflow loops showing above the rising flux rope, which are evident in EUV and X-ray images (\efig~1; \smv~1). The outflow loops grow higher with time, as an indication of continuous reconnection at increasing altitudes. These structures are intensively heated, containing dominant hot plasma with temperatures of 10--20 MK (Fig.~\ref{fig:f4_outflow}b). Hinode/EIS observation of the high-temperature spectral line Fe XXIV 192.03~\AA\ ($\sim$18~MK) also shows the cusp structures (Fig.~\ref{fig:f4_outflow}d,e; \efig~6; Methods). Particularly, the two cusps exhibit Doppler motions in opposite directions along the line of sight (LOS) from Earth: the one in the north is red-shifted and the other faint one in the south is blue-shifted. Comparing the EIS LOS velocity with the magnetic configuration of the AR, the Doppler red and blue shifts agree well with the bi-directional motions of two magnetic outflows produced by the external reconnection (Fig.~\ref{fig:f4_outflow}f, \efig~7). Spectral fitting shows that the outflow jet component exhibits excess line broadening with non-thermal velocities of $>$200 \kms, moving with a LOS Doppler velocity of $>$300 \kms\ (\efig~6; Methods). According to the AR geometry, the Doppler motion gives a reconnection outflow velocity of $>$500 \kms, which is comparable to the Alfv\'{e}n speed in the solar corona (Methods). If taking the rising flux rope, which participates in the external reconnection, as one magnetic inflow, and assuming the measured outflows approximately moving at the local Alfv\'{e}n speed, the non-dimensional reconnection rate, $M_{\rm A} \approx V_{\rm in} / V_{\rm out}$, is estimated as about 0.07 at around 21:21~UT (Fig.~\ref{fig:f4_outflow}c). Based on the height distribution of coronal Alfv\'{e}n speed $V_{\rm A}$, the estimated Alfv\'{e}n Mach number for the external reconnection above the flux rope, $M_{\rm A} = V_{\rm in} / V_{\rm A}$, varies from about 0.02 to 0.2 between $\sim$21:13 and 21:42~UT (Fig.~\ref{fig:f4_outflow}c). This reconnection rate is rather high, falling into the range of fast reconnection in solar flare cases \cite{Petschek1964,Lin2005,Gou2017}.

The eruption finally fails to escape, confined within $\sim$200~Mm ($<0.3~R_s$) above the solar limb. No substantial disruption is observed in the background corona during the flare, and the high-lying coronal loops above the AR remain closed (\efig~1, \efig~8), evidencing a confined flare (Methods). The hot flux rope structure fades from view during the flare gradual phase. Prominence material falls down to the solar surface during the eruption. In particular, a significant portion of the prominence mass does not trace the original flux rope structure as during the initiation phase but shows a distinct displacement. For example, some material is observed to move into the newly formed cusp above the flux rope; some threads fall along one leg of the reconnection outflow in the north; some fall back to the remote footpoint in P2 on the visible solar disk (Fig.~\ref{fig:f5_confined}a--c). IRIS high-resolution spectral and imaging observations on the northern leg of the prominence show complete Doppler red shifts from the falling threads after $\sim$21:22~UT, indicating a velocity of $\sim$200 \kms\ to the far side of the Sun (Fig.~\ref{fig:f5_confined}d; Methods). Considering that the prominence threads before the eruption trace magnetic field lines of the flux rope, the substantial displacement of prominence mass during the eruption indicates a transfer of magnetic flux from the rope into magnetic outflows, probably via the external reconnection occurring on the flux rope.

\section*{Discussion}

The question arises why the eruption failed even though the associated flare is intense. This event in the multipolar magnetic topology is notable for involving two reconnection processes during the eruption, i.e., standard flare reconnection below the magnetic flux rope and external reconnection above it (see the illustration in Fig.~\ref{fig:f6_cartoon}). The reconnection occurring at different locations may play different roles in the eruption process. 

Reconnection below the flux rope is a well-understood process in the standard model of solar eruptions and can facilitate the acceleration of the magnetic flux rope by supplying additional poloidal magnetic flux and reducing the tension of overlying fields \cite{Lin2000,Vrsnak2008}. This process is usually featured as the flare--CME coupling via the synchronization between the flare SXR (HXR) emission and the CME velocity (acceleration) \cite{Zhang2001,Temmer2010,Gou2020}. In this observation, multiple HXR bursts (e.g., $>$25~keV, coming from the dense flare loop region in the AR center; Methods) and microwave bursts are also associated with additional velocity rises during both episodes of the flare energy release (\efig~2), indicative of simultaneous high-energy particle acceleration and flux rope acceleration by fast reconnection at the flare CS. 

The external reconnection occurring above the flux rope may have an opposite effect compared to the flare reconnection underneath. While the breakout reconnection in the initiation phase opens the restraining field above the flux rope \cite{Antiochos1999,Karpen2012}, the subsequent external reconnection involving the magnetic flux rope itself can significantly reduce its magnetic flux and thus, on the contrary, cause deceleration. In the observation, the initial weak footpoint brightening in the remote polarity P2 is associated with significant acceleration of the flux rope (e.g., at $\sim$21:09 UT, $>$0.5~km/s$^2$; \efig~2e). However, when strong external reconnection occurs to the flux rope, a sharp velocity drop is observed (e.g., from $\sim$300 to 200~\kms\ at around 21:13 UT; \efig~2d). The velocity decrease can be interpreted in two aspects: on one hand, the external reconnection peels off the magnetic flux of the rope, causing an apparent drop of the rope front; on the other hand, the reduction of magnetic flux in the rope results in a decrease in the upward hoop force, thus decelerating the rope structure. The external reconnection between the flux rope and surrounding loops can significantly erode the flux rope structure, similar to the process in numerical simulations \cite{Aulanier2019,Jiang2023} and in observations occurring in the solar corona \cite{Gou2023} and during the CME propagation in interplanetary space \citep{Ruffenach2012}. In this event, although the overlying field is also partially opened by the external reconnection, the direct erosion on the flux rope itself makes a more significant impact on the eruption by reducing the hoop force, as evidenced by the pronounced velocity decrease.

In the observation, when intense external reconnection occurs on the flux rope, a strong episode of flare reconnection is also recorded as shown by the high-energy HXR peak immediately following the velocity decrease (\efig~2). This result suggests a close relationship between the two reconnection processes. On the other hand, strong flare reconnection facilitates the flux rope's acceleration, which further supplies faster inflow for the reconnection above. Thus, the flare reconnection below the rope and the external reconnection above it are coupled to each other. Meanwhile, they also compete with each other in accelerating the flux rope, which can significantly affect the eruption. As a result, the combined processes cause a continuous, dynamic variation as well as a replacement of magnetic flux in the flux rope \cite{Gou2023}. It is difficult to quantify the change of magnetic flux during the eruption in this event, but the final failure, the measured high reconnection rate for the external reconnection, as well as the prominence mass transfer to outflows, suggest a significant removal of magnetic flux in the erupting flux rope, which finally stops accelerating.

The multipolar AR also exhibits a strong confinement of overlying magnetic fields. The peripheral magnetic polarities P2/N2, where large-scale overlying fields are rooted, contain much more magnetic flux than the core polarities P1/N1, which host the erupting magnetic flux rope. The estimated flux ratio of P2/P1 is $>$2.5. We also examine the torus instability by calculating the decay index $n=-d\ln B_h/d\ln h$ as a function of the coronal height (Methods). The decay index profile above the prominence exhibits a saddle-like shape \cite{Luo2022} (Fig.~\ref{fig:f5_confined}e), indicating that the rising flux rope enters a local torus-stable regime (the critical height is as large as $\sim$180~Mm) soon after it becomes unstable at a low altitude ($\sim$20~Mm). The terminal height of the flux rope measured in off-limb EUV and X-ray observations is about 140~Mm, which falls in the stable region.

The unprecedented detail of observations in the event build a clear physical picture of the failed eruption in a multipolar magnetic system (Fig.~\ref{fig:f6_cartoon}). The results highlight the contrasting role of external reconnection to the traditional flare reconnection, i.e., flare reconnection underneath facilitates the flux rope's acceleration, whereas the above external reconnection on the rope induces deceleration by diminishing its flux. Simulations of the magnetic breakout model in multipolar configuration suggest that the reconnection between two lateral loops at the null point will strengthen the confinement of the central arcade \cite{DeVore2008,Lynch2008}, which can lead to a failed eruption. In this event, reconnection at the null point occurs on the central flux rope rather than between loops at two sides, which still reduces the restraining tension force above the flux rope. However, the final failure of the eruption suggests that other processes, such as erosion of the flux rope, play a crucial role. More generally, for flares within a bipolar magnetic configuration, external reconnection of the magnetic flux rope can also occur at the quasi-separatrix layers (QSLs) \cite{Demoulin1996} wrapping around the rope, which separate the twisted rope field lines from the untwisted overlying fields. Both observations and numerical simulations in the bipolar field topology have revealed the simultaneous occurrence of external reconnection on the flux rope and flare reconnection underneath during the solar eruption, mapped by the dynamic evolution of chromospheric flare ribbons \cite{Aulanier2019,Gou2023}. The two processes compete in altering the magnetic flux in the rope and will make a significant impact on the eruption. If the flare reconnection is strong enough and continuously supplies more flux into the rope, the rope will erupt successfully to form a large-scale CME \cite{Gou2023}. Otherwise, the flux rope will erode or even disintegrate during the eruption, leading to a failed eruption \cite{Ji2003,Jiang2023}. Therefore, the results provide broad implications for both eruptive and confined solar events in different magnetic topologies.

The detailed observations on the Sun provide insights into the study of stellar flares and eruptions beyond the solar system. The relation between solar flares and CMEs has shown a significant increase in the CME association with the flare class \cite{Yashiro2006}; however, large flares originating from solar ARs with a large magnetic flux tend to be confined in statistical results \cite{Li2020}. Our study of an intense yet non-eruptive flare provides insights into the detailed dynamic evolution and underlying mechanisms of failed CME eruption, by involving different reconnection processes in a complex region as well as strong magnetic confinement. In particular, stars can have significantly stronger magnetic fields and more complex magnetic topologies than the Sun, which imply a high rate of failed CME eruptions \cite{Donati2009,AlvaradoGomez2018}. The occurrence of failed eruptions on late-type stars adds another layer of difficulty in the search for stellar CMEs. Furthermore, the wealth of solar observations provides implications for stellar flare observations such as through the combination of spectroscopic measurement and multi-wavelength detection, including but not limited to UV, EUV, X-ray, and radio wavelengths. The comprehensive observation sheds light on the importance of multi-messenger astrophysics and astronomy.

\section*{Methods}

\subsection*{Multi-wavelength imaging}

The event was imaged by multiple spacecraft in multiple wavelengths. AIA \cite{Lemen2012} onboard SDO \cite{Pesnell2012} in geosynchronous orbit takes full-Sun images around the clock, capturing the eruption at the west solar limb. The AIA data are processed to level 1.5 after registering and co-aligning, which provides a spatial scale of 0.${''}$6 per pixel and a temporal cadence of 12s in seven EUV channels and of 24s in UV channels. During the flare, AIA automatically applied alternating long-short exposure times in multiple EUV channels to accommodate bright emissions. To remove the saturation of flare loops with long exposures, we make composite images by combining long and short exposure time pairs. We mainly use AIA 131~\AA\ (primarily from Fe XXI emission line, $\log T = 7.05$), 304~\AA\ (He II, $\log T = 4.7$), and 171~\AA\ (Fe IX, $\log T = 5.85$) channels, which feature the rising flux rope structure, prominence, and large-scale overlying loops (\efig~1; \smv~1). We also use AIA 1600~\AA\ (C IV) to study the footpoint brightening on the solar disk.

The AIA data in six EUV channels (except for 304~\AA) are used for the diagnosis of coronal plasma temperature and density. We perform the differential emission measure (DEM) analysis using the sparse inversion algorithm \cite{Cheung2015}, after deconvolving the point spread functions of AIA data in each channel. We calculate DEMs in the temperature range of $\log T$ = 5.5 -- 7.6 with an interval of $\Delta\log T$ = 0.05, derive the EM-weighted mean temperature $\langle T\rangle=\frac{\sum \mathrm{EM}(T_i) \times T_i}{\sum \mathrm{EM}(T_i)}$, and obtain the total $\mathrm{EM}=\sum \mathrm{EM}(T_i)$ in all temperature ranges. The coronal electron density $n_e$ is further calculated by solving $\mathrm{EM}=\int n_e^2dl$, assuming a fully ionized plasma along the LOS length $l$ (measured from SolO observation) in the optically thin corona.

Hinode/XRT \cite{Golub2007} observed the eruption with multiple filters including Be-thin, Be-med, Be-thick, and Al-thick. XRT images during the flare were taken with a high cadence and a pixel scale of $\sim 1''$, within a field of view (FOV) of $\sim 400''\times400''$. XRT images are calibrated to level 1 and co-aligned to AIA images. In the event, XRT is unique in imaging hot plasma emissions associated with the erupting magnetic flux rope and outflows, but excluding the cool, dense prominence material which is visible in all AIA EUV channels (\efig~1). 

At the time of the event, SolO \cite{Mueller2020} was located at a heliocentric distance of $\sim$0.31~AU from the Sun, separated by $\sim 44^\circ$ in longitude and 0.2$^\circ$ in latitude from Earth. The Full Sun Imager of the Extreme Ultraviolet Imager (EUI-FSI) \cite{Rochus2020} on SolO takes EUV images of the Sun in 174~\AA\ ($\sim$1~MK) and 304~\AA\ ($\sim$0.08~MK) channels, with a temporal cadence of 10~min and a spatial scale of $\sim 4.''4$ for this event. We use the level 2 EUI data. During the eruption, the FSI images were unfortunately blurred from 21:10 to 21:40 UT, thus we use the data before and afterward to study the flare brightening observed on the solar disk. The observing times of EUI images are corrected to account for the light travel time difference ($\sim$5.7 min) to SolO and Earth.

\subsection*{Confinedness of the event}

The event produces a failed prominence eruption and a confined solar flare. The flare is associated with a high class (above M9; note that it can be even higher since it is partially occulted behind the limb) as defined by the peak SXR flux of Geostationary Operational Environmental Satellites (GOES) and is accompanied by a long-duration gradual phase as in typical eruptive flares. However, this event is confined as demonstrated in various observations from multiple satellites and instruments. 

In EUV and X-ray observation of the low solar corona, high-cadence images taken by SDO/AIA and Hinode/XRT show that although the eruption is initiated, no substantial plasma or structure erupts outward to form a CME. The prominence and the associated flux rope structure are observed to be confined within $\sim$200~Mm ($<0.3~R_s$) above the solar limb. The large-scale overlying coronal loops in AIA cool channels remain closed during the whole process (\efig~1\&2a). The confinedness is further verified by GOES Solar Ultraviolet Imager (SUVI) with a larger FOV ($>1.6~R_s$; see \efig~8 and \smv~2). No significant disturbance in the background corona is observed during the flare (\efig~8h,i). No coronal dimming evidencing the expansion and/or evacuation of coronal plasma due to an eruptive CME \cite{Veronig2025} (see the difference image in \efig~8a and flare light curves in \efig~8d) or other signature of erupting CME plasma is found (\smv~2).

A diffusive disturbance is observed in white light (WL) in the coronagraph after the flare. The structure shows up as a bright front but dissipates quickly and becomes indiscernible beyond a distance of about 7 solar radii ($R_s$; see \efig~8b,c and \smv~2). Such a WL front tends to be classified as a CME event when it appears in the coronagraph. However, high-cadence EUV observations demonstrate that the flare does not produce substantial eruption, and no other eruption is recorded in nearby regions. To study the nature of the WL disturbance and its possible association with the flare, we examine the kinematic evolution in images taken by the Large Angle and Spectrometric Coronagraph (LASCO) \cite{Brueckner1995} onboard the Solar and Heliospheric Observatory (SOHO), along the same direction as the height measurements obtained from SDO/AIA and GOES/SUVI. The derived velocity shows that the WL structure propagates at $\sim$200--300~\kms~and decelerates significantly after $\sim5~R_s$ (\efig~8e). The deceleration seems unrealistic if taking it as a slow CME, which would instead accelerate due to the drag force exerted by the ambient solar wind \cite{Cargill2004,Vrsnak2007}. Moreover, the velocity of the WL structure does not align with that of the flux rope in the low corona, where the latter reaches a higher peak velocity much earlier and eventually decreases to zero (\efig~8e). The velocity behaviors suggest that the WL front is not a resultant CME from the flare but an independent feature. In the LASCO coronagraph, the structure quickly evolves into an irregular morphology (\smv~2), suggestive of the lack of its coherence and of a substantial driver. There is no corresponding structure detected in situ by other spacecraft. Based on the kinematic evolution, the morphological change, and the confined behavior in the low corona, we argue that the WL structure is not a coherent CME associated with the flare. It is probably a wave-like disturbance associated with the impulsive energy release during the flare and failed eruption \cite{Warmuth2015,Howard2016,Morosan2023}.

\subsection*{Flare spectroscopy}

Hinode/EIS \cite{Culhane2007} is a scanning slit spectrometer that operates in the two wavelength channels 170--212~\AA\ and 246--292~\AA. It performed 35 consecutive raster scans between 18:10~UT and 23:22~UT with the study \textsf{FlareResponse01}, which observes an extensive list of spectral lines covering a broad temperature range. Each raster has an FOV of $240''\times340''$ obtained through 80 steps of 3$''$ with the 2$''$ slit. The exposure time is 5~s, the exposure cadence is 6.6~s, and the raster cadence is 8~m 55~s. The plate scale is 1$''$ in the $y$ dimension and 22~m\AA\ in the spectral dimension. For this article we focus on the EIS spectral line Fe XXIV 192.03~\AA\ (with a formation temperature of $\log T = 7.25$, $\sim$18~MK) obtained in raster 22, between 21:17 and 21:26~UT. We fit a Gaussian to the line profiles to obtain maps of spectral line intensity, Doppler velocity, and non-thermal velocity (Fig.~\ref{fig:f4_outflow}d,e; \efig~6a--c). The post-flare loops at low altitudes are saturated, while the two cusp-shaped outflows are well highlighted in the Doppler and non-thermal velocity maps. 

The Fe XXIV line shows complex profiles in and around the cusp regions and so multi-Gaussian fits were performed in selected regions to better constrain the Doppler motions. \efig~6(a) shows five locations, labeled 1--5, with four in the northern cusp and one in the southern cusp. The line profiles for the five locations are shown in \efig~6d--h, where two and three Gaussian fits are displayed. The Fe XII 192.39~\AA\ line ($\log T = 6.1$, $\sim$1~MK) is found in the same spectral window and is included in the fit. Since this line is emitted from the background corona then it is assumed to have zero Doppler shift and thus provides a wavelength calibration for the Fe XXIV 192.03~\AA\ line. Locations 1--3 and 5 clearly show multiple components to the Fe XXIV line, with one component with a high speed and the other at a lower speed. Location 4 near the tip of the cusp only shows a single, high-speed component. The data clearly show that the northern cusp exhibits large redshifts, whereas the southern cusp shows large blueshifts. These flows are consistent with the reconnection outflows shown schematically in Figures~\ref{fig:f4_outflow} and \ref{fig:f6_cartoon}. The Gaussian components also show large non-thermal broadening, up to 220~\kms\ at location 4, and Fe XXIV emission extends out to 600~\kms\ at locations 2--4. We estimate the outflow velocity using the LOS Doppler velocity from spectral fitting, $V_{\rm out}=V_{\rm Doppler}/\cos(\theta)$, where $\theta$ is measured as $\sim 50^\circ$ in the magnetogram if assuming the outflows move perpendicularly to the PIL (see the schematic plot in Fig.~\ref{fig:f4_outflow}f). The outflow velocity is typically about 500~\kms\ (up to 900~\kms\ at the red wing) for the selected locations (Fig.~\ref{fig:f4_outflow}c), comparable to the Alfv\'{e}n speed in the solar corona. 

During the event, IRIS \cite{DePontieu2014} ran coordinated observations with Hinode as part of IRIS/Hinode Observing Program (IHOP) 409. IRIS focused on the northern part of the eruption near the limb, providing simultaneous spectra and context images with a high spatial resolution of 0.17$''$ (Fig.~\ref{fig:f5_confined}d). IRIS SJI images were taken in the 1330~\AA\ channel with a cadence of 10~s in a $60''\times 60''$ FOV. IRIS spectra were taken in a medium coarse 8-step raster with a spatial step of 2$''$ and a temporal cadence of 9.6~s per step. IRIS spectral observations show complex line profiles from the prominence material. Spectra including C II 1334/1335~\AA, Si IV 1403 \AA, and Mg II k 2796~\AA, contain both blue and red shifted components at different locations during the prominence eruption. When the prominence threads fall along the leg after $\sim$21:22~UT, the spectral profiles are completely red-shifted (Fig.~\ref{fig:f5_confined}d), showing a LOS Doppler velocity of $\sim$50 -- 80 \kms. The redshifts indicate the prominence mass falling to the far side of the Sun, probably along the outflow leg in P2. Combining the plane-of-sky velocity measured in IRIS SJI 1330~\AA\ images, the prominence threads fall at a speed of $\sim$200~\kms\ along the leg, implying a small LOS angle of $\sim10-20^\circ$.

\subsection*{Flare HXR emissions}

STIX \cite{Krucker2020} on SolO recorded HXR emissions from accelerated particles during the flare. The bottom row of STIX pixels was partially occulted during this event, thus we only use data from the top row pixels to generate the lightcurve and reconstruct HXR images. For this double-peak flare, STIX pixel data have a temporal resolution of 1s from 20:58 to 21:14 UT during the first energy-release episode and of 2s from 21:14 to 21:43 UT around the second peak, in an energy range of 4 -- 150 keV. The light travel time difference is corrected to compare with observations near Earth (with a time delay of $\sim$5.7 min on Earth). STIX HXR images are reconstructed with a 20~s duration primarily using the CLEAN algorithm, also compared with the maximum entropy (MEM\_GE) and expectation maximization methods to check the reliability \cite{Massa2023}.

HXR emissions during the flare show accelerated electrons with energies up to 100~keV (\efig~2f). The HXR images during the eruption show source emissions located in the center of the AR, near the magnetic polarities P1 and N1 (Fig.~\ref{fig:f3_rec}b). We did not find STIX HXR sources in the remote polarities such as in P2 and N2, probably due to the dynamic range of the instrument and the high density and/or strong magnetic strength near the main flare region. Around the strongest HXR peak at $\sim$21:13~UT, HXR images in high energy ranges (e.g., 36 -- 50, 50 --100 keV) show two main sources, from nonthermal footpoint emissions by the thick-target bremsstrahlung; the low-energy range (6 -- 10 keV) shows a bright thermal source probably from the flare loop top (Fig.~\ref{fig:f3_rec}b).

\subsection*{Microwave observation}

Microwave observations were obtained using EOVSA \cite{Gary2018}, a solar-dedicated radio interferometric array that provides critical diagnostics of coronal magnetic fields and energetic electrons through broadband imaging spectroscopy. The data were processed using EOVSA’s standard calibration pipeline, followed by self-calibration to improve image fidelity. Imaging was performed at a temporal cadence of 1~s, with the frequency coverage spanning 1 -- 18 GHz, divided into 50 spectral windows each with a bandwidth of approximately 325 MHz. The resulting spatial resolution, or synthesized beam size, varies inversely with frequency and is given by approximately $140''/{\nu}_{\rm GHz}$.

During the flare, the microwave emission is dominated by intense gyrosynchrotron continuum emission, superimposed with two distinct, coherent bursts at frequencies of $\sim$2--6 GHz, each lasting over short time intervals, approximately three seconds (\efig~5). The first burst exhibits a frequency drift from 3.5 to 6.1 GHz over 3 seconds around 21:13:11 UT, while the second one drifts from 2.8 to 4.8 GHz over 4 seconds near 21:13:16 UT. The microwave sources are spatially distributed along a predominantly horizontal, east-west direction at different coronal heights above flare loops, except for during the two coherent bursts. Notably, the peak emission of the two coherent structures is systematically displaced toward the north, coinciding with the vicinity of the magnetic null point, suggestive of reconnection-driven electron acceleration and energy release. The low-to-high frequency drift of the microwave emissions above the flux rope indicates accelerated electrons moving into a region with a higher density or a stronger magnetic field, probably associated with the dense prominence and magnetic flux rope.

\subsection*{Magnetic configuration}

To investigate the magnetic structure associated with the eruption, we study the magnetic configuration of the solar AR. The Helioseismic and Magnetic Imager (HMI) \cite{Scherrer2012} on SDO and the Full Disc Telescope of the Polarimetric and Helioseismic Imager (PHI-FDT) \cite{Solanki2020} on SolO make measurements of the photospheric magnetic field. The event occurred at the solar limb from SDO's viewpoint, so only HMI magnetic data taken in a few days before (three days at least) can be used. SolO observed the event on the solar disk and PHI-FDT imaged the photospheric magnetogram on the same day. Comparing the magnetic field observed in previous days from SolO/PHI (\efig~4a--d), one can see that the AR undergoes rapid and substantial evolution, particularly significant emergence of negative magnetic flux between the two major positive ones. Thus we use PHI magnetic data observed on the day of the flare to investigate the pre-flare configuration.

We use level 2 SolO/PHI-FDT data, which provides the LOS magnetogram on the photosphere with a spatial scale of $\sim 3.''6$ ($\sim$0.8~Mm). We use the most recent magnetogram prior to the flare, which was taken at 04:40~UT. The AR was centered at about S12W40 in heliographic Stonyhurst coordinates as seen from SolO, which is still close to the solar disk center. To investigate the magnetic topology, we extrapolate the coronal magnetic fields using the PHI photospheric LOS magnetic data of the AR as the boundary. We re-project the magnetogram into heliographic Carrington coordinates (\efig~4), and rebin the magnetic data by a factor of 2 before modeling. We extrapolate the potential (current-free) magnetic field in the corona based on the Fourier transformation method \cite{Alissandrakis1981}, and model the magnetic field vertically at different heights with the same resolution as in the horizontal direction. 

The AR exhibits a complex multipolar magnetic field configuration with four major magnetic polarities distributed longitudinally (Fig.~\ref{fig:f2_AR}, \efig~4). We found an X-type magnetic null point in the corona, located at $\sim$46~Mm above the photosphere (\efig~4). The magnetic components in the y and z directions vanish at the X-point, associated with a guide field in the x direction. The magnetic field line below the X-point mainly connects from P1 to N1 in the core of the AR, but the orientation changes significantly above it to connect from P2 to N2 (\efig~4f). The X-type null is high-lying in the corona, at the intersection of magnetic separatrices for different magnetic flux systems, where magnetic reconnection will inevitably occur.

Based on the potential coronal field extrapolated from the photospheric magnetogram, we quantify the decay of horizontal magnetic field ($B_h$) with height ($h$) using the decay index, $n=-d\ln B_h/d\ln h$. The torus instability model predicts that a magnetic flux rope becomes unstable if it reaches a critical height $h\rm_{cr}$ where $n\rm_{cr}=1.5$ \cite{Kliem2006}, i.e., the magnitude of overlying field decreases sufficiently fast. We examine the decay index as a function of height above the central PIL in the AR, where the prominence and flux rope is observed in the initial phase. The mean decay index near the center of the eruptive structure exhibits a saddle-like profile (Fig.~\ref{fig:f5_confined}e), indicative of the presence of a torus-stable regime below $h\rm_{cr2}\approx$179~Mm shortly after the flux rope becomes unstable at $h\rm_{cr1}\approx$21~Mm. For this off-limb eruption as seen from Earth, the height of the flux rope and prominence can be directly measured in EUV and X-ray images with the least projection effect (\efig~2c), which are found to finally stop accelerating below $h\rm_{cr2}$. The results suggest that with the external reconnection involved, the flare reconnection during the eruption fails to accelerate the magnetic flux rope into a sufficiently large height (e.g., $>$180~Mm) to enable the torus instability set in.

We also calculate the coronal Alfv\'{e}n speed, $v_{\rm A}=\frac{B}{\sqrt{\mu_0 \rho}}$, using the extrapolated coronal magnetic field and plasma density estimated from the DEM analysis ($\sim 2 - 5 \times10^9$~cm$^{-3}$). The Alfv\'{e}n speed decreases from about 1500 to a few hundreds \kms\ above the height of magnetic X-point within the AIA's FOV, which is used to derive the Alfv\'{e}n Mach number $M_{\rm A}$ for the reconnection above the flux rope (Fig.~\ref{fig:f4_outflow}c).

\subsection*{3D reconstruction}

Taking advantage of the multi-viewpoint observation, we are able to reconstruct the 3D geometry of the coronal eruptive structure to compare with the magnetic field topology. We use a semi-circular shape to model the prominence/flux rope and the post-flare loop, adjusting its position, size, and azimuth on the Sun to best match the observation.

During the initiation of the eruption, the eruptive structure is simultaneously observed in the 304~\AA\ channels of SDO/AIA and SolO/EUI-FSI at 21:06~UT, as well as in 304~\AA\ of the Extreme Ultraviolet Imager (EUVI) \cite{Wuelser2004} on STEREO-A at 21:05~UT (\efig~3). STEREO-A/EUVI shows a similar view as SDO/AIA due to a small separation from Earth ($\sim10^\circ$), but the erupting prominence is still observed within the limb (in spite of the saturation), which helps consolidate the loop model by adding a third viewpoint. We adjust the loop orientation and height to match with EUV observations, and make sure the two footpoints anchored in opposite magnetic polarities in the SolO/PHI magnetogram (see the middle row in \efig~3). To compare with the magnetic configuration of the AR, we plot the reconstructed 3D structure together with the coronal magnetic field lines (Fig.~\ref{fig:f2_AR}a,b). The result nicely shows the eruptive picture in the initiation phase.

During the impulsive eruption, the EUV images by SolO/EUI-FSI and STEREO-A/EUVI are heavily saturated, so it is hard to identify the observational features. We use SDO/AIA in combination with the PHI-FDT and STIX HXR observations from SolO to reconstruct the front of the erupting flux rope, as well as the post-flare loop at a low altitude, observed at 21:12:42~UT (Fig.~\ref{fig:f3_rec}a,b). We similarly model a semi-circular loop shape which is anchored in opposite magnetic polarities. For the low-altitude flare loop, we make the footpoints of the 3D loop on two HXR sources in STIX 36 -- 50 keV image and the loop top co-spatial with the thermal source in 6 -- 10 keV. The two modeled loops show a height of $\sim$48~Mm and 9~Mm at the apex, respectively, both of which are highly sheared with respect to the PIL. Comparing with the coronal magnetic configuration, the front of the flux rope approaches the height of the X-point in the corona (Fig.~\ref{fig:f3_rec}a,b; \efig~7). We note that the magnetic X-point is obtained from the pre-eruption background coronal magnetic field, and its location will likely change during the eruption. Nevertheless, the magnetic topology as well as the 3D loop geometry nicely indicates that magnetic reconnection occurs above the erupting flux rope, as evidenced in various observations.

\section*{Supplementary Information}

\subsection*{Supplementary Video 1}
SDO/AIA multi-wavelength observation of the failed solar eruption on 2024 March 30. The blended images are AIA 131~\AA\ (cyan, $\sim$10~MK), 304~\AA\ (red, $\sim$50,000~K), and 171~\AA\ (yellow, $\sim$1~MK) observations, which feature the hot flux rope and outflow loops, cool prominence, and warm overlying loops high above, respectively.

\subsection*{Supplementary Video 2}
GOES/SUVI and SOHO/LASCO observations covering 6 hours since the flare initiation. The SUVI images have an extended FOV of $\sim1.6~R_s$. The 195~\AA\ base difference images on the left show that no substantial structure escapes from the flare region at the west limb, and the overlying loops high above the AR remain closed. The LASCO C2 (2 -- 6~$R_s$) and C3 ($>6~R_s$) images taken the closest in time with SUVI 195~\AA\ are plotted on the right. A diffusive front appears in the running difference images after 21:36 UT above the west limb, but it quickly becomes irregular and dissipates in the outer corona. For comparison, another successful eruption near the northeast limb causes prominent coronal dimmings and significant disturbances near the Sun, and produces a CME that is clearly visible in the coronagraph.

\subsection*{Data Availability}
The data used in the study are publicly available for download from the mission archives and can be accessed by searching for the observing instrument and the date of the event. The SDO/AIA data can be downloaded from the Joint Science Operations Center at \url{http://jsoc.stanford.edu/}, under the series \texttt{aia.lev1\_euv\_12s} and \texttt{aia.lev1\_uv\_24s}. Data from EUI, PHI, and STIX are available from the Solar Orbiter Archive at \url{https://soar.esac.esa.int/soar/}. Data from XRT and EIS are available from the Hinode Data Archive at \url{https://hinode.msfc.nasa.gov/data_archive.html}. IRIS data are publicly available at \url{https://iris.lmsal.com/data.html}.

\subsection*{Acknowledgments}
The authors acknowledge the data from SDO, Hinode, IRIS, Solar Orbiter, STEREO, and SOHO. The authors also thank David Williams for the SolO/PHI data. T.G. and K.K.R acknowledge the support by contract 8100002705 from Lockheed-Martin to SAO. P.R.Y. acknowledges funding from the Hinode project and the GSFC Internal Scientist Funding Model competitive work package program. A.M.V. acknowledges the Austrian Science Fund (FWF) projects no. 10.55776/I4555 and 10.55776/PAT7894023. B.Z. acknowledges NASA grant 80NSSC23K1057 and NSF grant AGS-2301382. S.Y. and B.C. acknowledge NASA grant 80NSSC24K1242 to NJIT.
Solar Orbiter is a space mission of international collaboration between ESA and NASA, operated by ESA. Hinode is a Japanese mission developed and launched by ISAS/JAXA, with NAOJ as domestic partner and NASA and STFC (UK) as international partners. It is operated by these agencies in co-operation with ESA and NSC (Norway). IRIS is a NASA small explorer mission developed and operated by LMSAL with mission operations executed at NASA Ames Research Center and major contributions to downlink communications funded by ESA and the Norwegian Space Centre. The EOVSA was designed, built, and is now operated by NJIT as a community facility. EOVSA operations are supported by NSF grant AGS-2436999 and NASA grant 80NSSC20K0026 to NJIT.

\subsection*{Author Contributions}
T.G. conceptualized and led the study, performed the analysis and led the writing. K.K.R., A.M.V., and B.Z. contributed to the interpretation of data and results. P.R.Y. performed the EIS data analysis and contributed to the interpretation. X.C., S.Y., and B.C. contributed to the EOVSA data analysis and interpretation. All authors discussed the results and contributed to the manuscript text.

\subsection*{Competing Interests}
The authors declare no competing interests.

\subsection*{Corresponding Authors}
Correspondence to Tingyu Gou (\url{tingyu.gou@cfa.harvard.edu})

%%%%%%%%%%%%%%%%%%%%%%%%%%%%%%%%%%%%%%%%% figures
\clearpage
\renewcommand{\figurename}{\textbf{Fig.}}
\setcounter{figure}{0}
\spacing{1}

\begin{figure}[htbp]
	\centering
	\includegraphics[width=\textwidth]{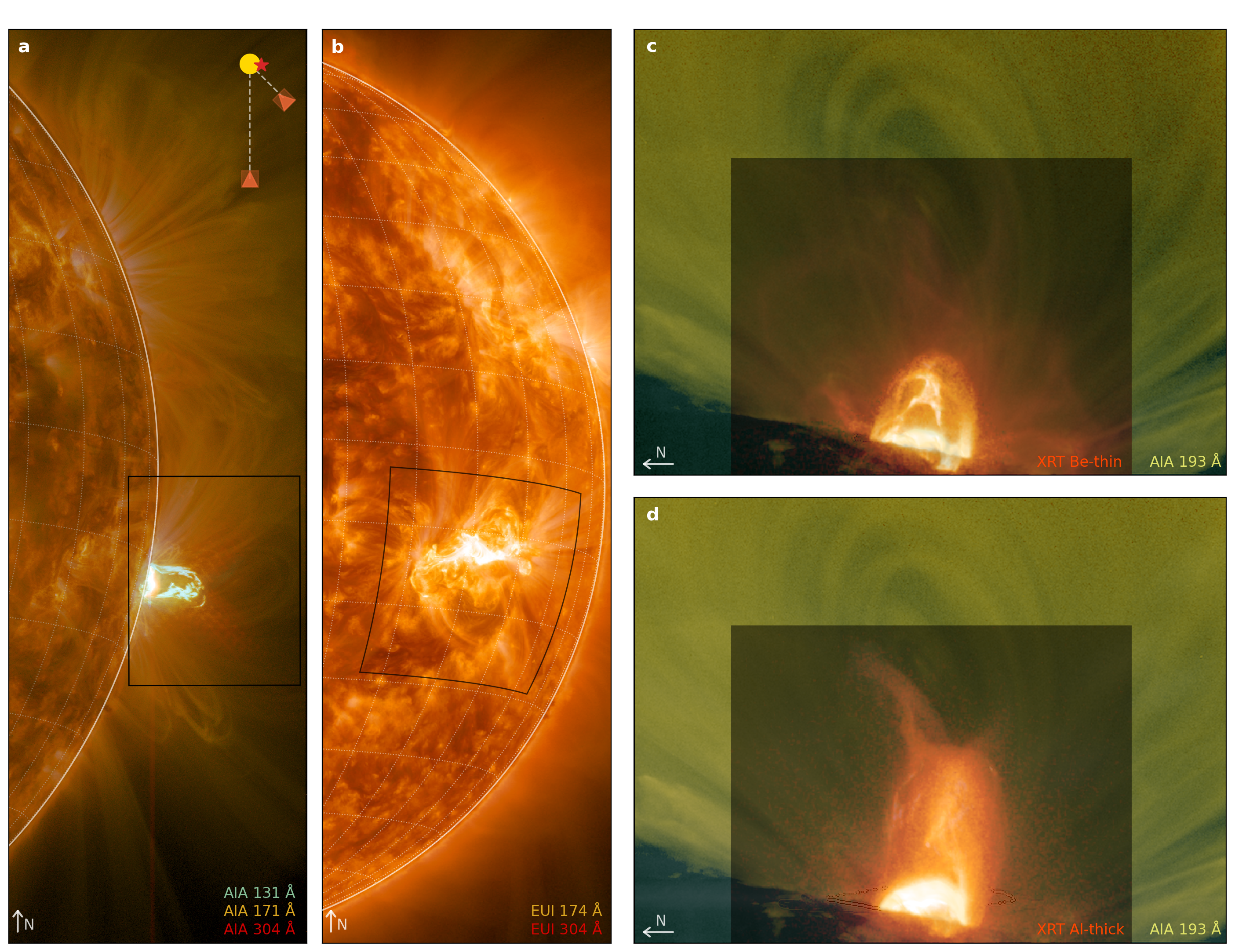}
    \caption{\small \textbf{Multi-viewpoint and multi-wavelength observation of the failed solar eruption on 2024 March 30.} \textbf{a}, The prominence eruption at the west solar limb observed at 21:20~UT in AIA 131~\AA\ (cyan), 171~\AA\ (yellow), and 304~\AA\ (red) channels of SDO/AIA near Earth. The inserted plot at the top right shows locations of the Sun (yellow circle), the prominence (red star), Earth and Solar Orbiter (orange triangles) in the ecliptic plane. \textbf{b}, The same event on the solar disk observed at 22:00~UT in 174~\AA\ (yellow) and 304~\AA\ (red) of EUI FSI on Solar Orbiter, which is $\sim 44^\circ$ separated from Earth in longitude. \textbf{c,d}, Close view of the event observed by SDO/AIA and Hinode/XRT at 21:12~UT and 21:20~UT during the flare (box region in a, rotated 90$^\circ$ anticlockwise). The XRT image (red) features high-temperature plasmas ($\geq$10~MK) associated with the magnetic flux rope and cusp-shaped outflows above. The AIA 193~\AA\ image (yellow) features cooler overlying loops ($\sim$2~MK) at high altitudes. The arrow at the bottom left indicates the direction of solar north.}
    \label{fig:f1_overview}
\end{figure}

\begin{figure}[htbp]
	\centering
	\includegraphics[width=\textwidth]{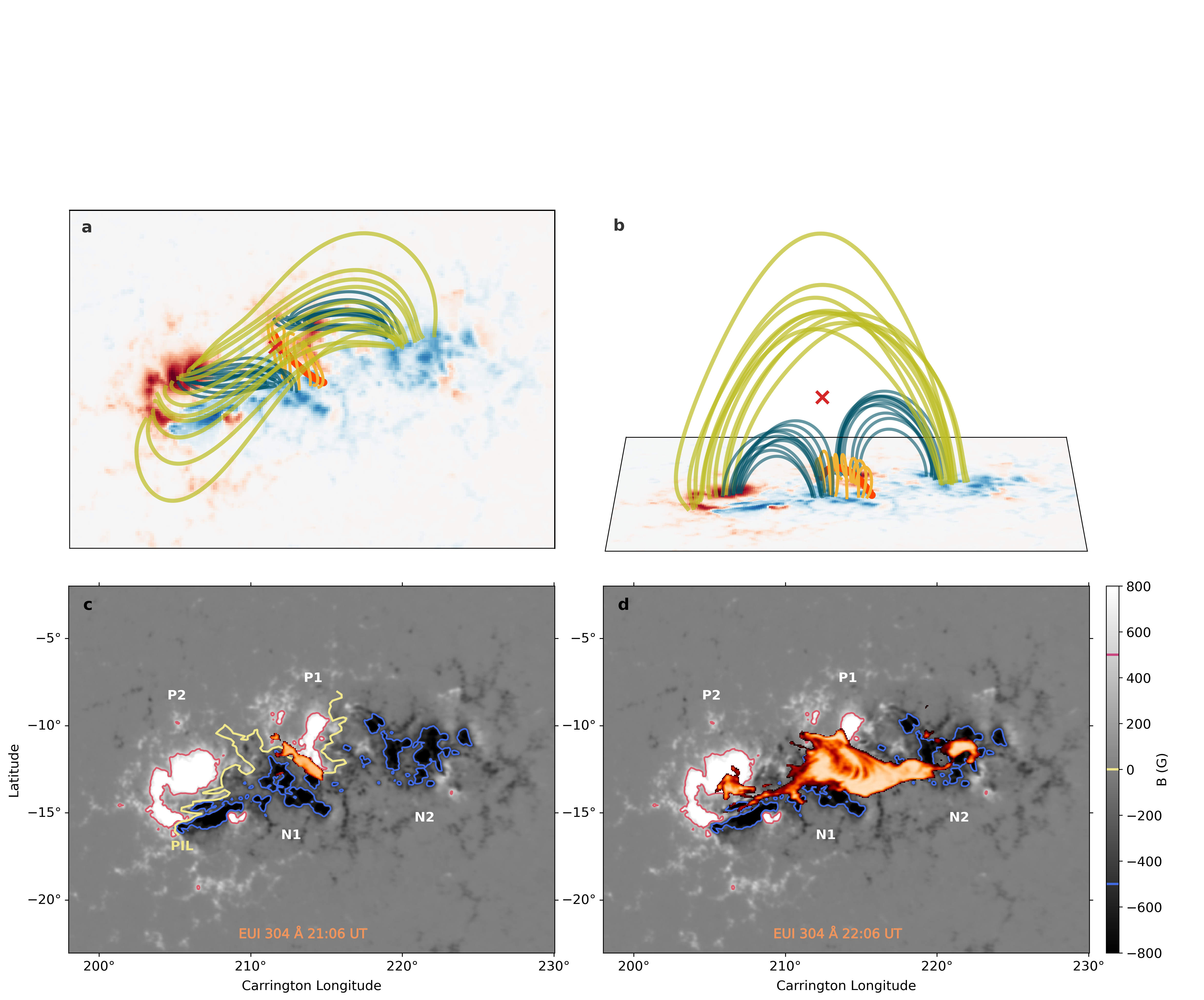}
    \caption{\small \textbf{Magnetic configuration of the solar AR.} \textbf{a,b}, Top and face views of coronal magnetic field lines on top of the photospheric magnetogram (red-blue scale; same as the gray-scale image in c,d). The yellow, dark blue, and light orange curves show representative magnetic field lines connecting different magnetic polarities in the multipolar AR, obtained from a potential coronal field extrapolation (Methods). The prominence observed in EUV images is plotted with the thick orange curve (see also the middle row in \efig~3; Methods), which rises up from the core of the AR embedded inside the overlying fields. The location of the coronal magnetic X-point is marked as a red cross symbol, which is $\sim$46~Mm above the photosphere, in between the field lines connecting P1/N1 and P2/N2 (see also \efig~4). \textbf{c,d}, EUV brightenings (red-orange) before and after the eruption, plotted on the pre-flare magnetogram (gray scale). The four major magnetic polarities are labeled as P1, P2, N1, and N2, as outlined by red and blue contours with magnetic field strengths $B=\pm$500~G, and separated by a highly curved PIL ($B=0$; yellow) in the AR. The prominence in the initiation phase is observed lying along the central segment of the PIL.}
    \label{fig:f2_AR}
\end{figure}

\begin{figure}[htbp]
	\centering
	\includegraphics[width=\textwidth]{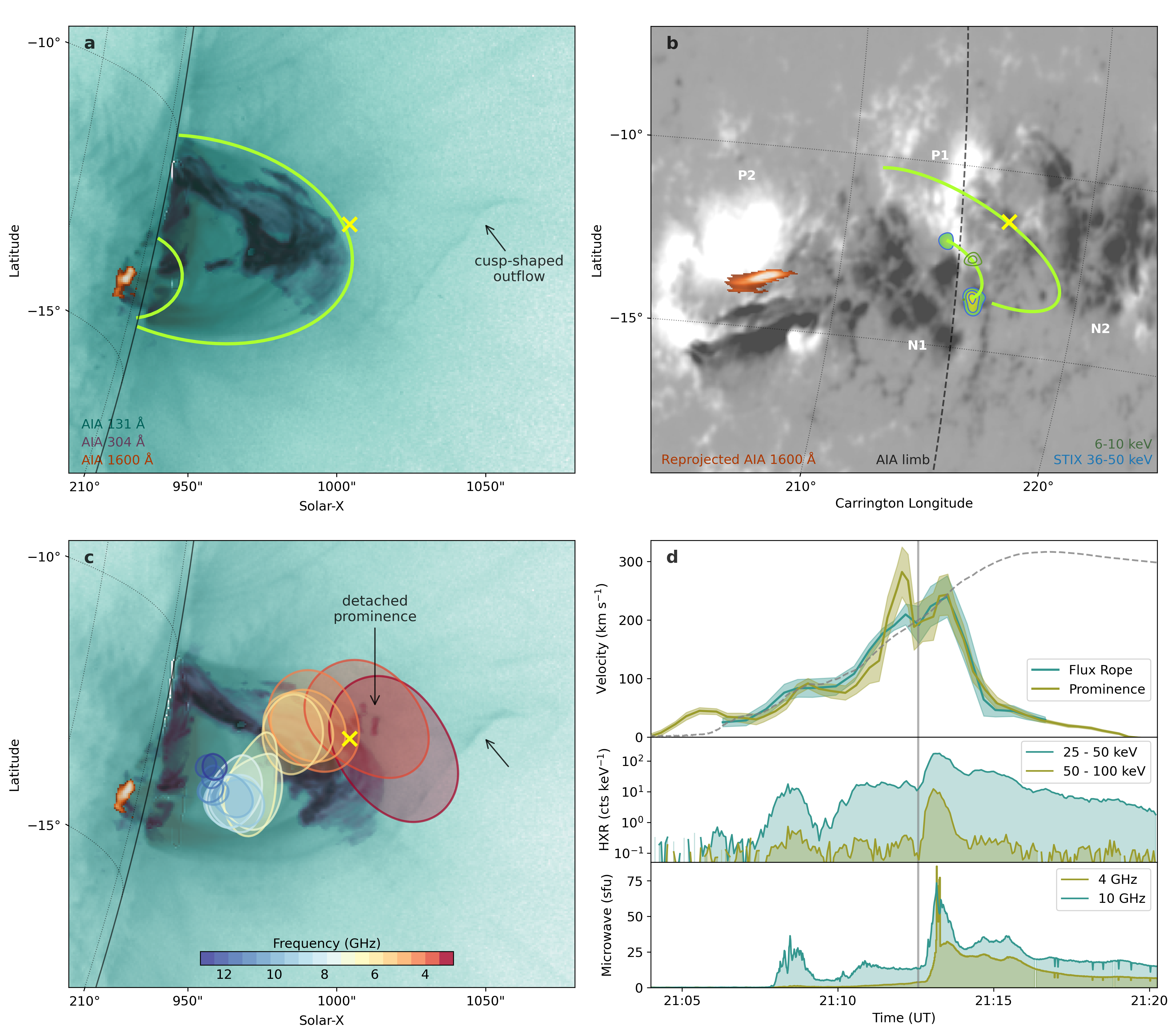}
    \caption{\small \textbf{Observational evidence for magnetic reconnection during the flare.} \textbf{a}, AIA 131~\AA\ (dark cyan) and 304~\AA\ (dark purple) blended images at 21:12:42~UT, overlaid with simultaneous footpoint brightening in AIA 1600~\AA\ on the solar disk (orange). A pronounced cusp-shaped outflow, labeled by a black arrow, appears above the rising flux rope in AIA 131~\AA. \textbf{b}, Pre-flare magnetogram overlaid with re-projected AIA 1600~\AA\ footpoint brightening, and STIX HXR sources from the low-altitude post-flare loops observed at $\sim$21:13~UT. STIX 36 -- 50 keV sources at 50, 70, and 90\% contour levels are plotted in blue, showing nonthermal HXR emissions from two footpoints; 6 -- 10 keV source contours at 70, 90\% show thermal emissions from the loop top. The two semi-circles in a and b model a low-altitude post-flare loop and the front of the erupting flux rope, respectively, seen from different viewpoints (Methods). \textbf{c}, Similar background AIA image as in panel a but taken at 21:13:30~UT. The filled colored contours show EOVSA microwave sources above 80\% of the maximum intensity at different frequencies, observed at 21:13:11~UT. The yellow cross symbol in a -- c indicates the location of the magnetic X-point in the corona (see also Fig.~\ref{fig:f2_AR}). \textbf{d}, Temporal evolution of the rising velocity, the flare HXR and microwave fluxes. The velocity uncertainties are derived through error propagation of the height measurement uncertainties (see also \efig~2d). The GOES 1--8~\AA\ SXR flux is plotted with a dashed grey curve. The vertical line marks the time of a local minimum velocity at 21:12:35~UT.}
    \label{fig:f3_rec}
\end{figure}

\begin{figure}[htbp]
	\centering
	\includegraphics[width=\textwidth]{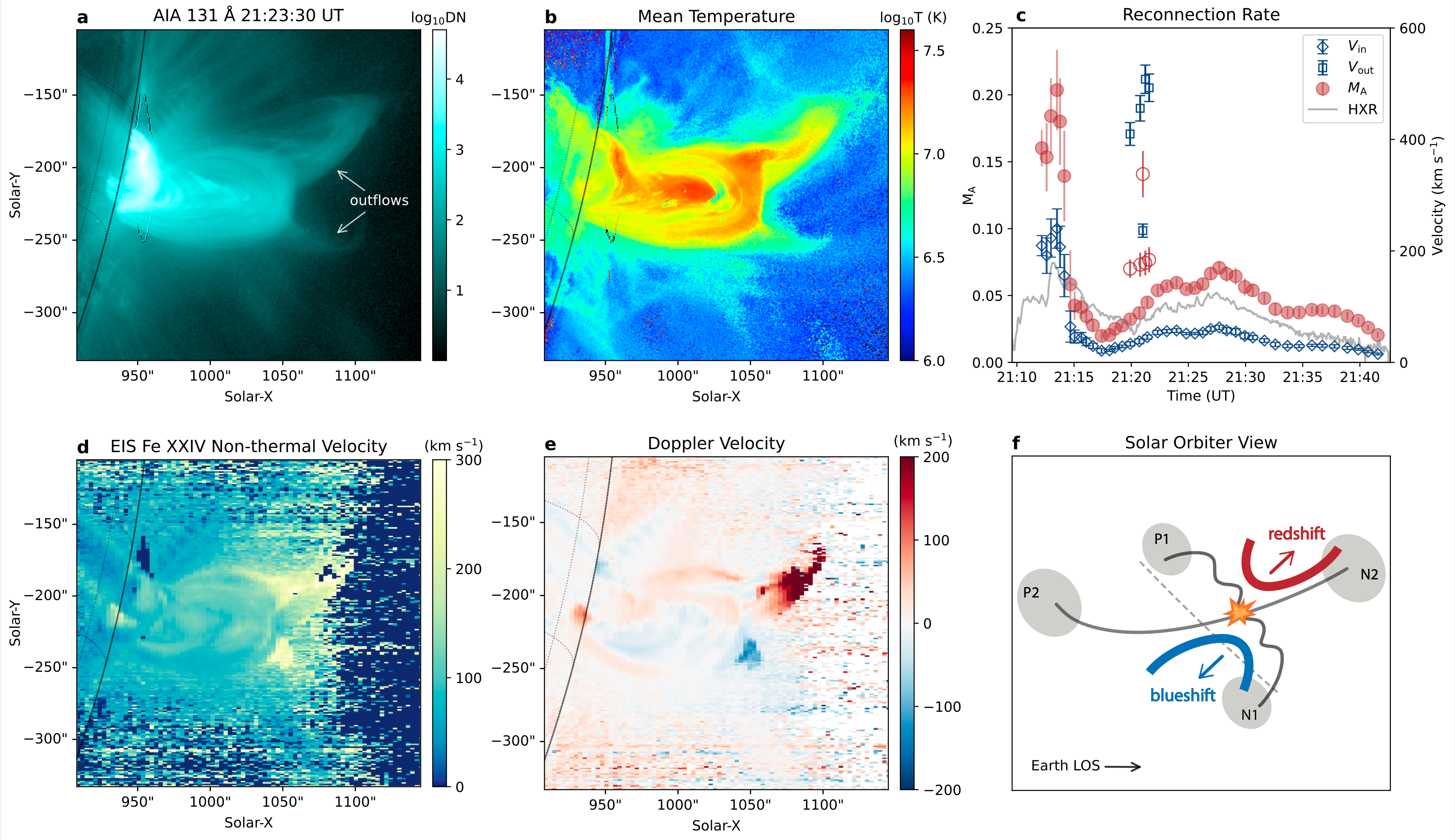}
    \caption{\small \textbf{Spectroscopic and imaging observations of reconnection outflows.} \textbf{a\&b}, AIA 131~\AA\ image and mean temperature map obtained from the DEM analysis (Methods), showing two hot, cusp-shaped outflows above the flux rope. \textbf{c}, Temporal evolution of inflow/outflow velocities, and the Alfv\'{e}n Mach number derived from the coronal Alfv\'{e}n speed ($M_{\rm A}=V_{\rm in}/V_{\rm A}$; Methods). The errors are derived through propagation of uncertainty. The five measurements of outflow velocities are derived from EIS Fe XXIV 192.03~\AA\ LOS Doppler velocity in the cusp regions (\efig~6; Methods), for which the estimated $M_{\rm A} \approx V_{in} / V_{out}$ are shown as hollow circular symbols. The gray curve shows the STIX 25 -- 50 keV HXR flux, plotted in an arbitrary axis. \textbf{d\&e}, Maps of non-thermal velocity and Doppler velocity from EIS Fe XXIV 192.03~\AA\ ($\sim$18~MK) during the raster between 21:17 and 21:26~UT (Methods), featuring two cusp structures with high non-thermal velocity and strong Doppler motions in opposite directions. \textbf{f}, Schematic sketch showing the two reconnection outflows in the north and south from SolO's viewpoint, associated with Doppler red and blue shifts as seen from Earth, respectively.}
    \label{fig:f4_outflow}
\end{figure}

\begin{figure}[htbp]
	\centering
	\includegraphics[width=\textwidth]{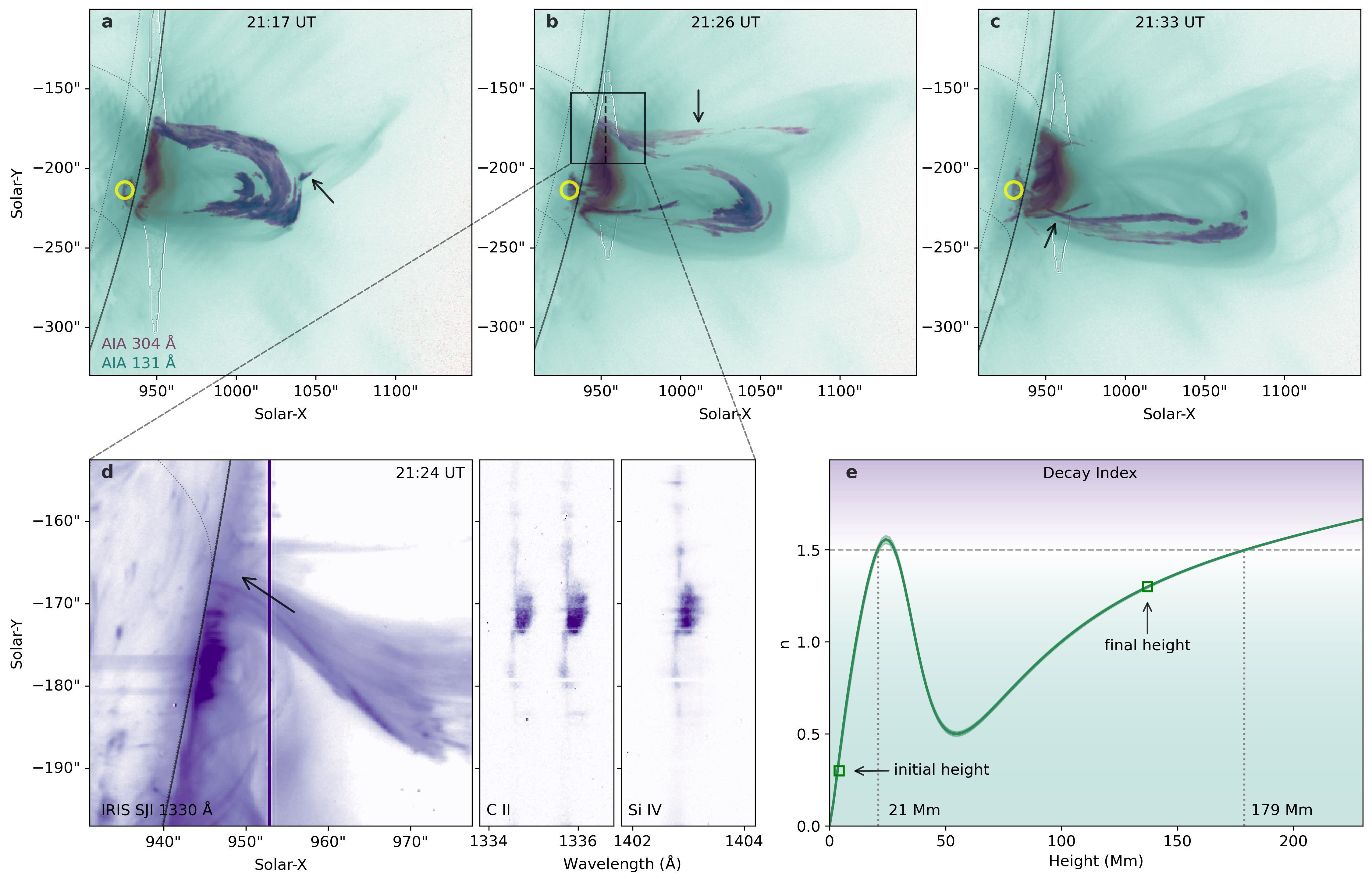}
    \caption{\small \textbf{Failure of the eruption after initiation.} \textbf{a--c}, AIA 131~\AA\ and 304~\AA\ blended images featuring the hot plasma structures and cool prominence material, respectively. The prominence mass moving into hot outflow loops and falling to the remote footpoint in P2 (yellow circle) are indicated by black arrows. \textbf{d}, IRIS SJI 1330~\AA\ image, and spectra of C II lines at $\sim$1335~\AA\ and Si IV 1403~\AA\ emission line from the slit on the prominence's leg. The falling prominence thread moves across the IRIS slit (indicated by the black arrow) and shows complete Doppler redshifts in the spectra. \textbf{e}, Decay index distribution with height above the photospheric PIL in the core of the AR (Methods). Light green colors in the background represent the torus-stable regime in general ($n<1.5$) while purple is unstable ($n>1.5$). The two values of the critical height are labeled and marked by two vertical dotted lines. The initial and terminal heights of the flux rope observed in EUV images are indicated as square symbols.}
    \label{fig:f5_confined}
\end{figure}

\clearpage
\begin{figure}[hbp]
	\centering
	\includegraphics[width=\textwidth]{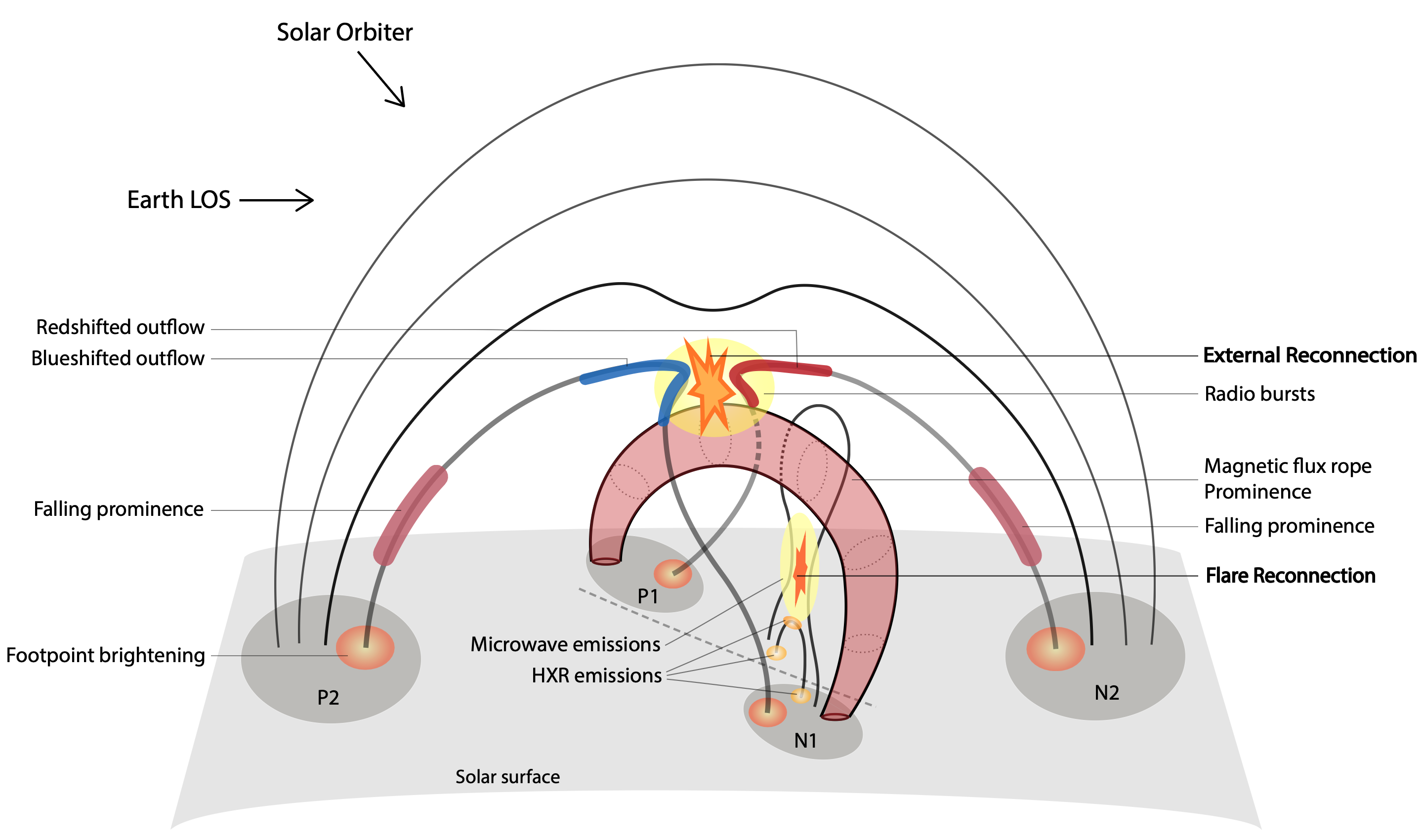}
    \caption{\small \textbf{Schematic illustration of the failed eruption in a multipolar magnetic topology.} A wealth of observational features are illustrated, evidenced by various instruments on multiple satellites. The magnetic flux rope rises in the core of the magnetic system. During the rise, magnetic reconnection occurs not only at the flare CS below the flux rope but also on its front. Footpoint brightenings and ribbons are observed in all four polarities, especially the one in P2 which is recorded as intense brightening when external reconnection occurs on the flux rope. HXR emissions are seen from the post-flare loops near the central PIL. Microwave observations show both gyrosynchrotron emission from the flare CS and intense radio bursts from the top of the flux rope. A pair of cusp-shaped, high-temperature outflows exhibits Doppler blue and red shifts in the spectrograph, respectively. Prominence mass is observed to transfer into the outflow loops and fall to the remote polarities. The complex AR is also associated with strong magnetic confinement from the large-scale overlying fields.}
    \label{fig:f6_cartoon}
\end{figure}

%%%%%%%%%%%%%%%%%%%%%%%%%%%%%%%%%%%%%%%%% extended data figures
\clearpage
\renewcommand{\figurename}{\textbf{Extended Data Fig.}}
\setcounter{figure}{0}
\spacing{1}

\begin{figure}[htbp]
	\centering
	\includegraphics[width=\textwidth]{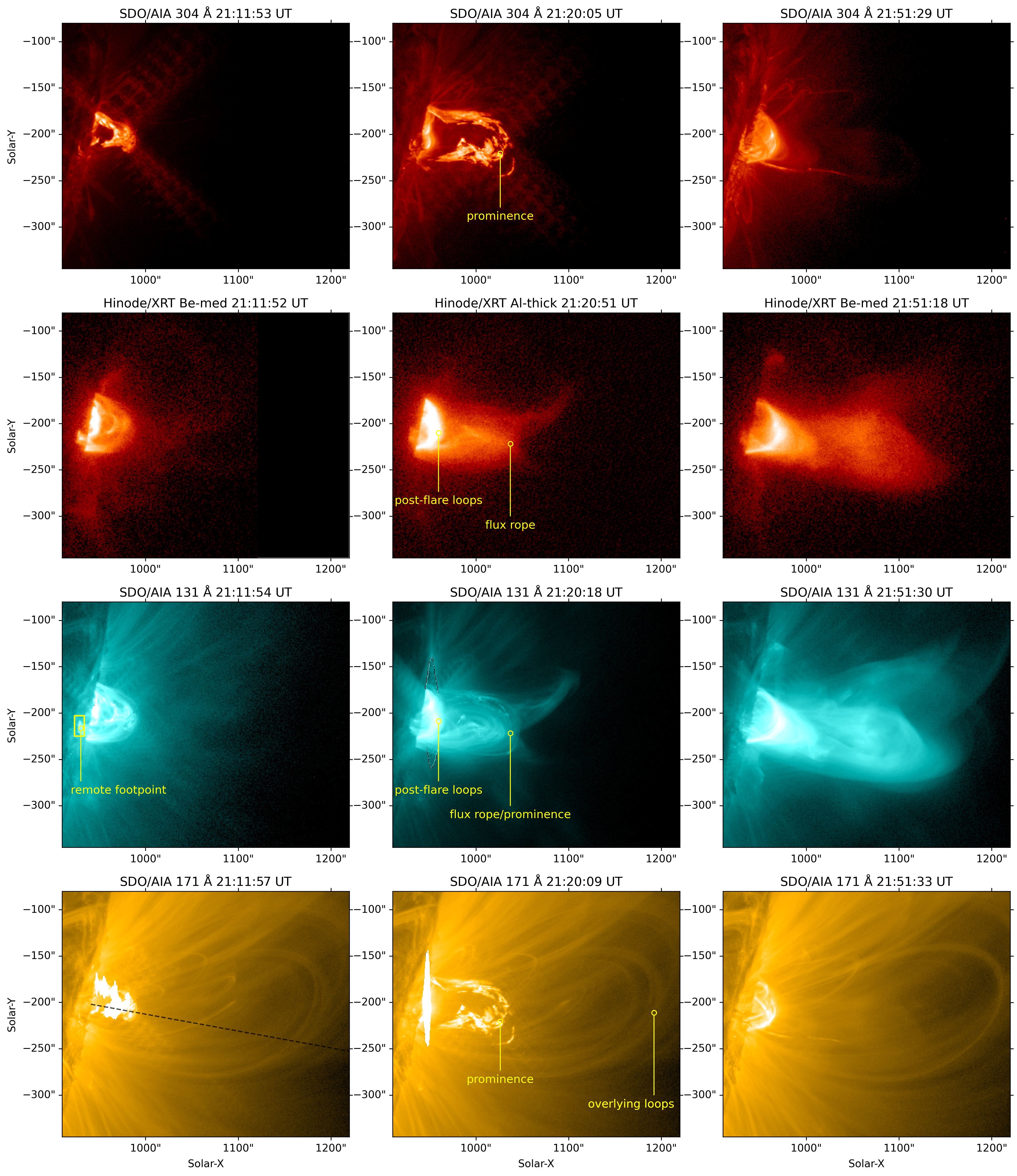}
    \caption{\small \textbf{EUV and X-ray observations of the failed solar eruption on 2024 March 30.} Images from top to bottom show AIA 304~\AA\ ($\sim$50,000~K), XRT Be-med/Al-thick ($\sim$10~MK), AIA 131~\AA\ ($\sim$10~MK) and 171~\AA\ ($\sim$1~MK) observations. The dashed line in the bottom left panel indicates the location of a virtual slit (starting at the solar limb) used to generate the height-time plots in Extended Data Fig.~2a\&b. The small rectangle in AIA 131~\AA\ image indicates the region to obtain light curve of the remote footpoint in AIA 1600 \AA, as shown in purple in Extended Data Fig.~2e.}
\end{figure}

\begin{figure}[htbp]
	\centering
	\includegraphics[width=0.8\textwidth]{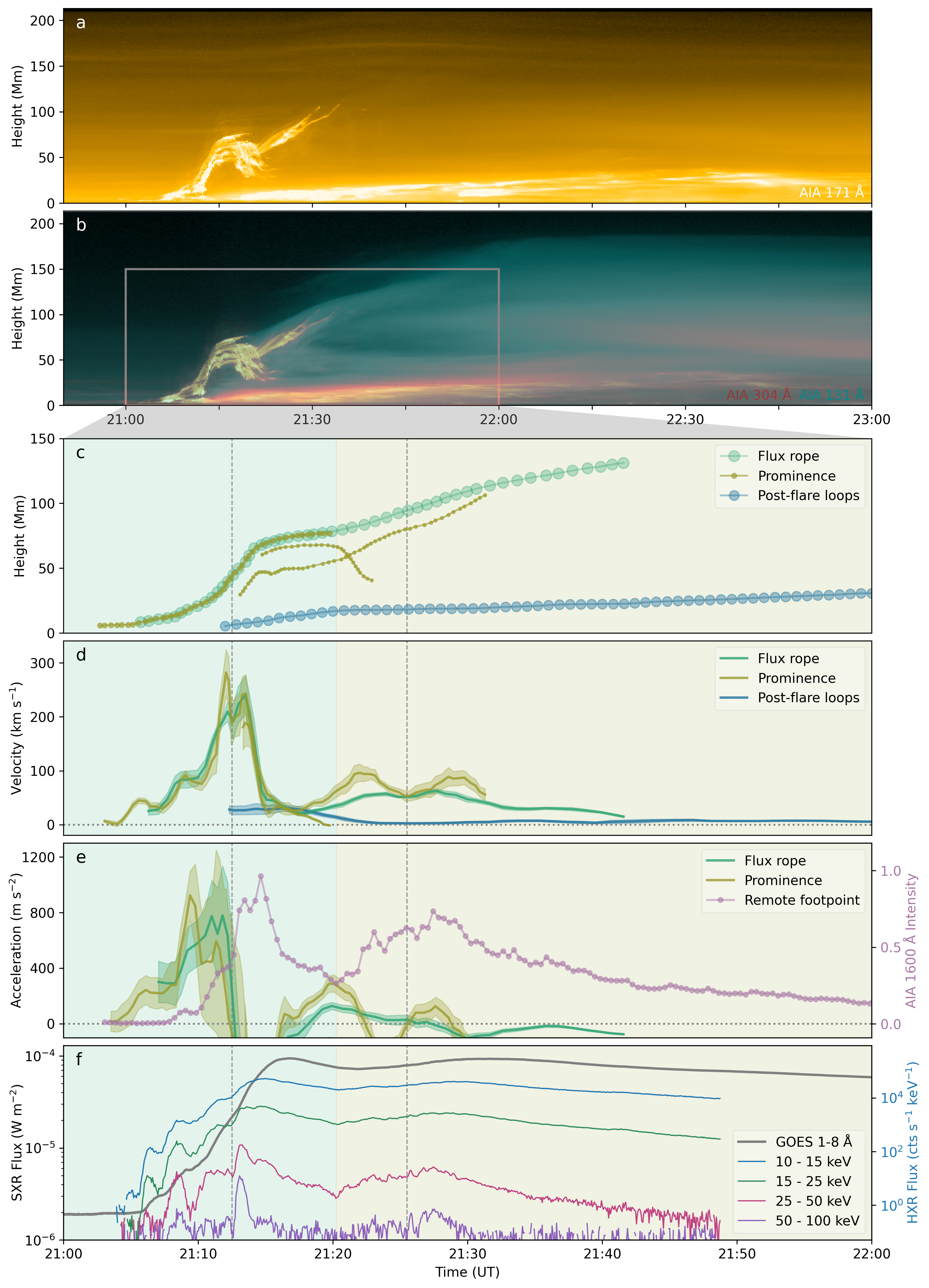}
    \caption{\small \textbf{Kinematics and timeline of the event.} a\&b, Height-time evolution as seen in the AIA 171~\AA, 131~\AA\ and 304~\AA\ channels through a virtual slit above the solar limb (the dashed line in Extended Data Fig.~1). c--e, Temporal evolution of the height, velocity, and acceleration of representative prominence threads (measured in AIA 304~\AA), the leading font of the magnetic flux rope (from AIA 131~\AA), and post-flare loops at low altitudes (from AIA 94~\AA). The uncertainties in velocity and acceleration are derived through error propagation of the height measurement uncertainties, which are assumed to be 5 pixels. The normalized footpoint brightness in AIA 1600~\AA\ images in the remote magnetic polarity P2 (see also Fig.~\ref{fig:f3_rec}) is plotted in purple in e. f, GOES and STIX X-ray flux of the flare. The two colored shadow regions in c--f indicate two episodes of the flare energy release. The two vertical dashed lines mark two remarkable local minima in the rising velocity of the erupting prominence (in d), which are associated with immediate increases in flare HXR emissions (e.g., $>$25 keV).}
\end{figure}

\begin{figure}[htbp]
	\centering
	\includegraphics[width=\textwidth]{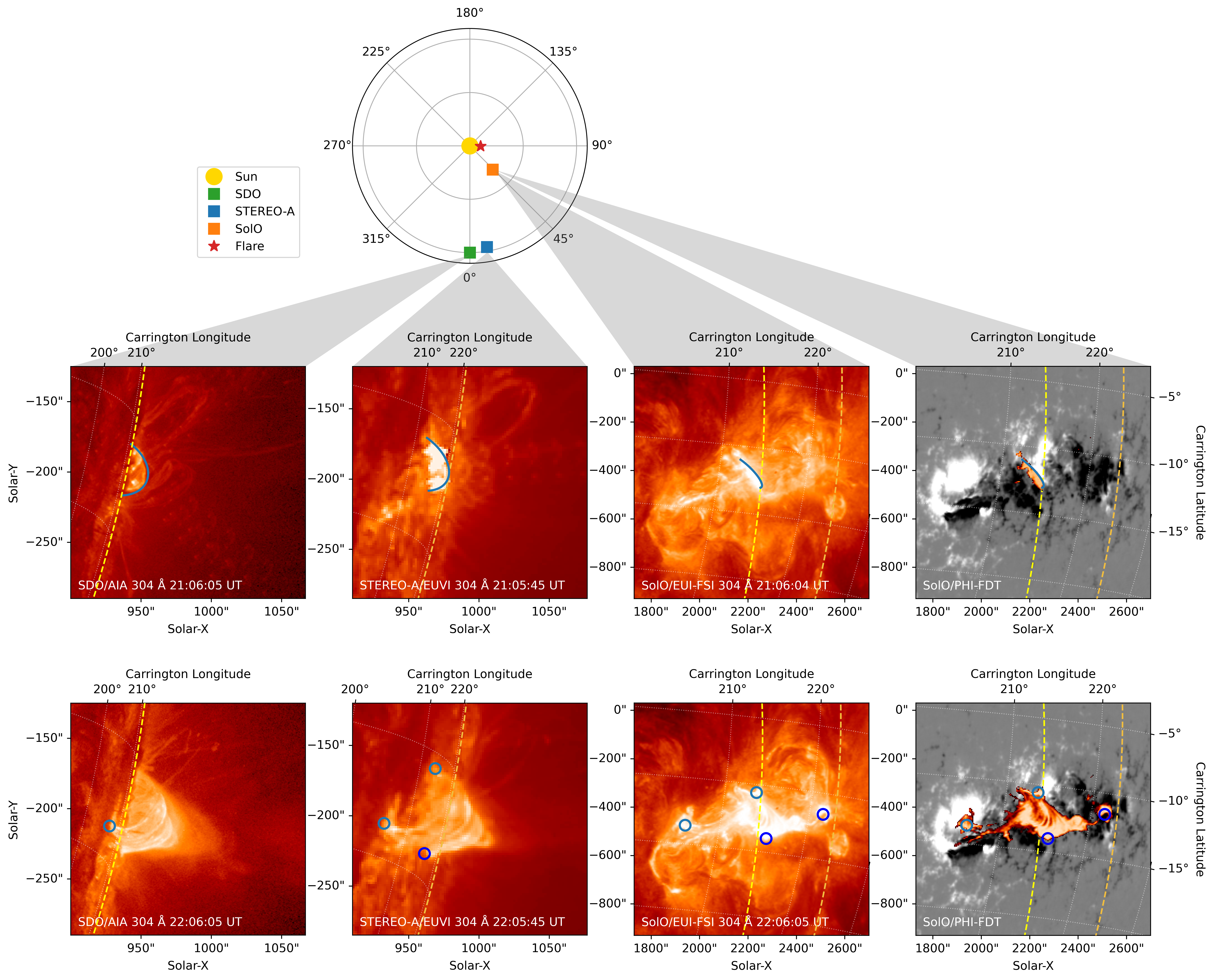}
    \caption{\small \textbf{Multi-viewpoint observation from multiple spacecraft.} The top panel shows the locations of SDO, STEREO-A, and SolO relative to the Sun in the ecliptic plane. STEREO-A (SolO) has a longitudinal separation from Earth by $\sim$10$^\circ$ (44$^\circ$). The two bottom rows show 304~\AA\ EUV images before and after the eruption, respectively, from SDO/AIA, STEREO-A/EUVI, and SolO/EUI, as well as EUI brightening on top of the SolO/PHI pre-flare magnetogram. The two dashed curves indicate locations of the solar limb seen in AIA and EUVI, respectively. The thick blue curve in the middle row shows the 3D model of the prominence from the multi-viewpoint observation (Methods). The circles in the bottom row mark the flare ribbon brightening observed in the four major magnetic polarities.}
\end{figure}

\begin{figure}[htbp]
	\centering
	\includegraphics[width=\textwidth]{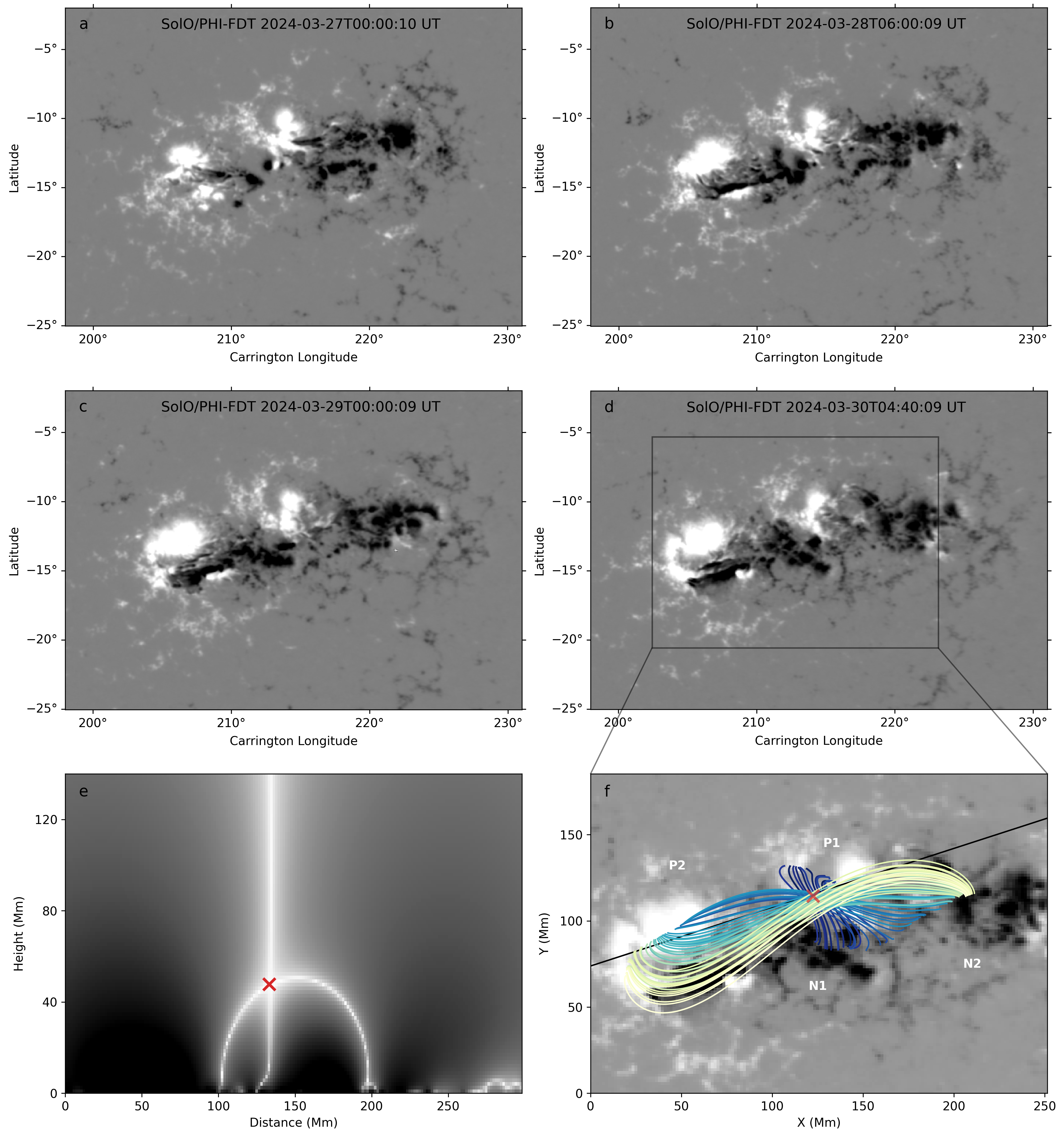}
    \caption{\small \textbf{Evolution and magnetic configuration of the solar AR.} a--d, Photospheric LOS magnetogram observed by SolO/PHI-FDT on different days prior to the event. The magnetic data are re-projected to the heliographic Carrington coordinates, scaled within $\pm$~800~G. e\&f, Magnetic topology of the AR from a potential magnetic field extrapolation (Methods). The height distribution of radial magnetic field magnitude ($|B_z|$, logarithmically scaled within 0--200~G) above the photosphere in e is made from a cut across four magnetic polarities, indicated by the black line in f. The X-type null point in the corona is marked as a cross symbol. The colored curves in f show representative magnetic field lines traced around the null point at different coronal heights.}
\end{figure}

\begin{figure}[htbp]
	\centering
	\includegraphics[width=0.96\textwidth]{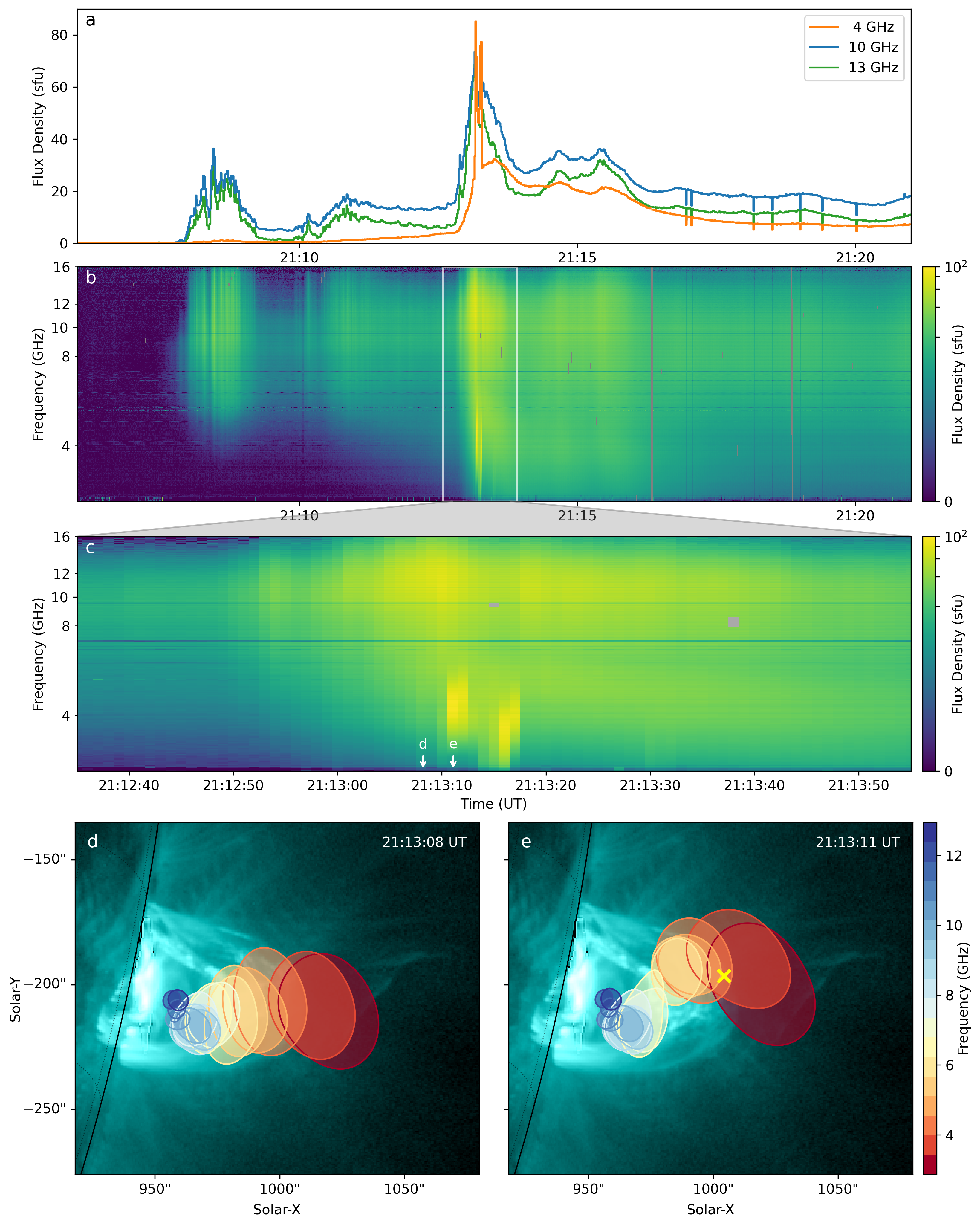}
    \caption{\small \textbf{Microwave observation from EOVSA.} a--c, EOVSA light curves at representative frequencies and the background-subtracted dynamic spectra. The zoom-in plot at around 21:13~UT in c highlights the two short-lived radio bursts on top of the flare gyrosynchrotron continuum emissions. d\&e, EOVSA microwave images in different frequencies before and during the first intense radio burst, overplotted on an AIA 131~\AA\ image observed near 21:13:12~UT as the background. The filled contours show microwave sources above 80\% of the maximum intensities from 2.9 to 12.9 GHz. The yellow cross in e indicates the location of the magnetic null point (see also Fig.~\ref{fig:f3_rec}c).}
\end{figure}

\begin{figure}[htbp]
	\centering
	\includegraphics[width=\textwidth]{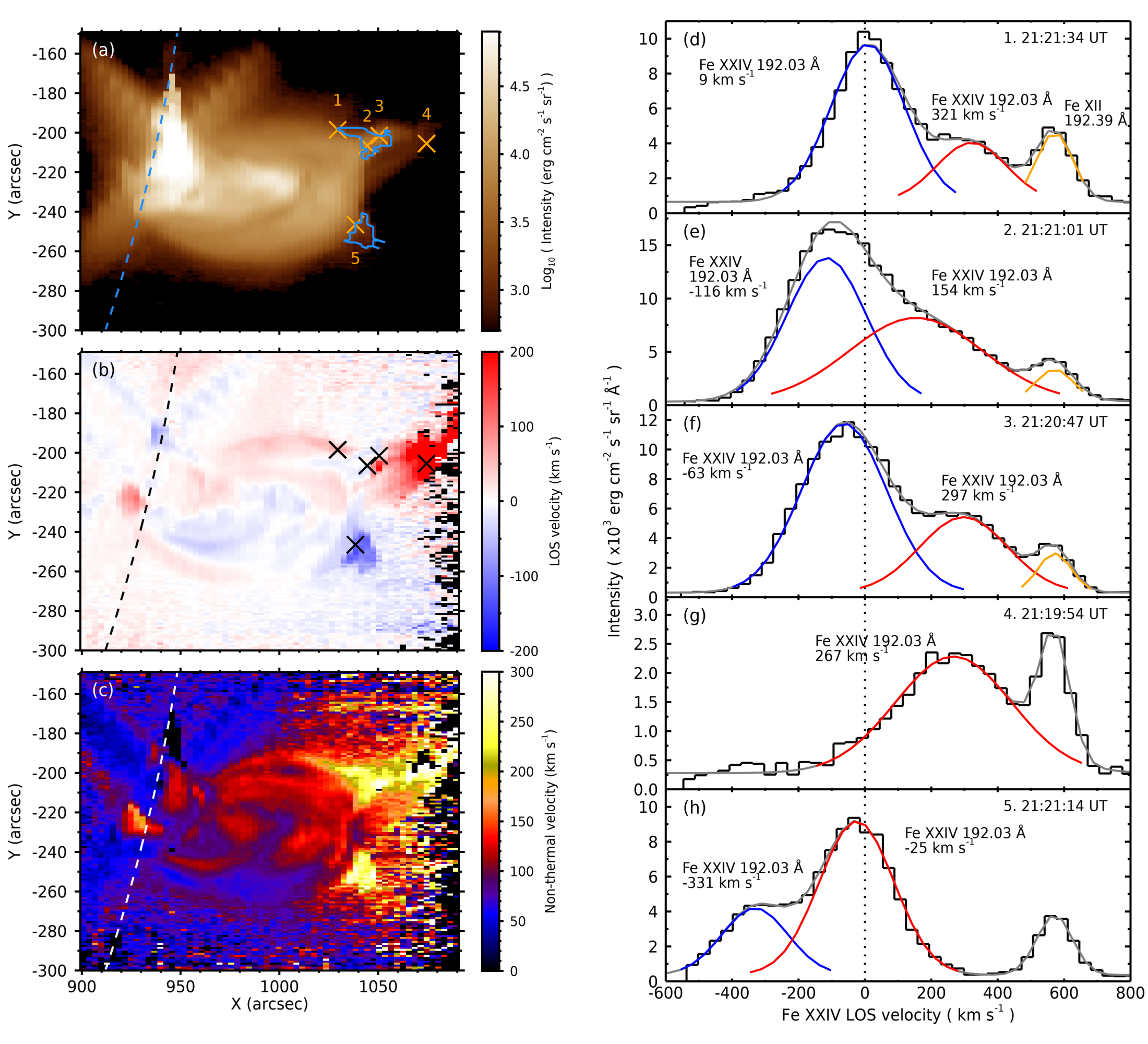}
    \caption{\small \textbf{Spectroscopic observation by Hinode/EIS.} a--c, maps of the Fe XXIV 192.03 \AA\ spectral line intensity, Doppler velocity, and non-thermal velocity, obtained from a Gaussian fit to the line profile (Methods). The dashed curves indicate the solar limb. The non-thermal velocities larger than 220 \kms\ in the two cusp regions are contoured by blue curves in the intensity map in (a). d--h, multi-Gaussian fits to spectral line profiles from five selected locations, which are indicated in a and b by cross symbols, labeled as numbers 1 -- 5. The blue and red curves show two Gaussian components fitted to the Fe XXIV 192.03 \AA\ line emission (18~MK), whose Doppler velocities are also indicated. The orange curve shows a fit to the Fe XII 192.39~\AA\ line (1~MK; peaking at around 600 \kms) coming from the background corona, and the gray curve shows a total fit to the line profile.}
    \label{efig:eis}
\end{figure}

\begin{figure}[htbp]
	\centering
	\includegraphics[width=0.8\textwidth]{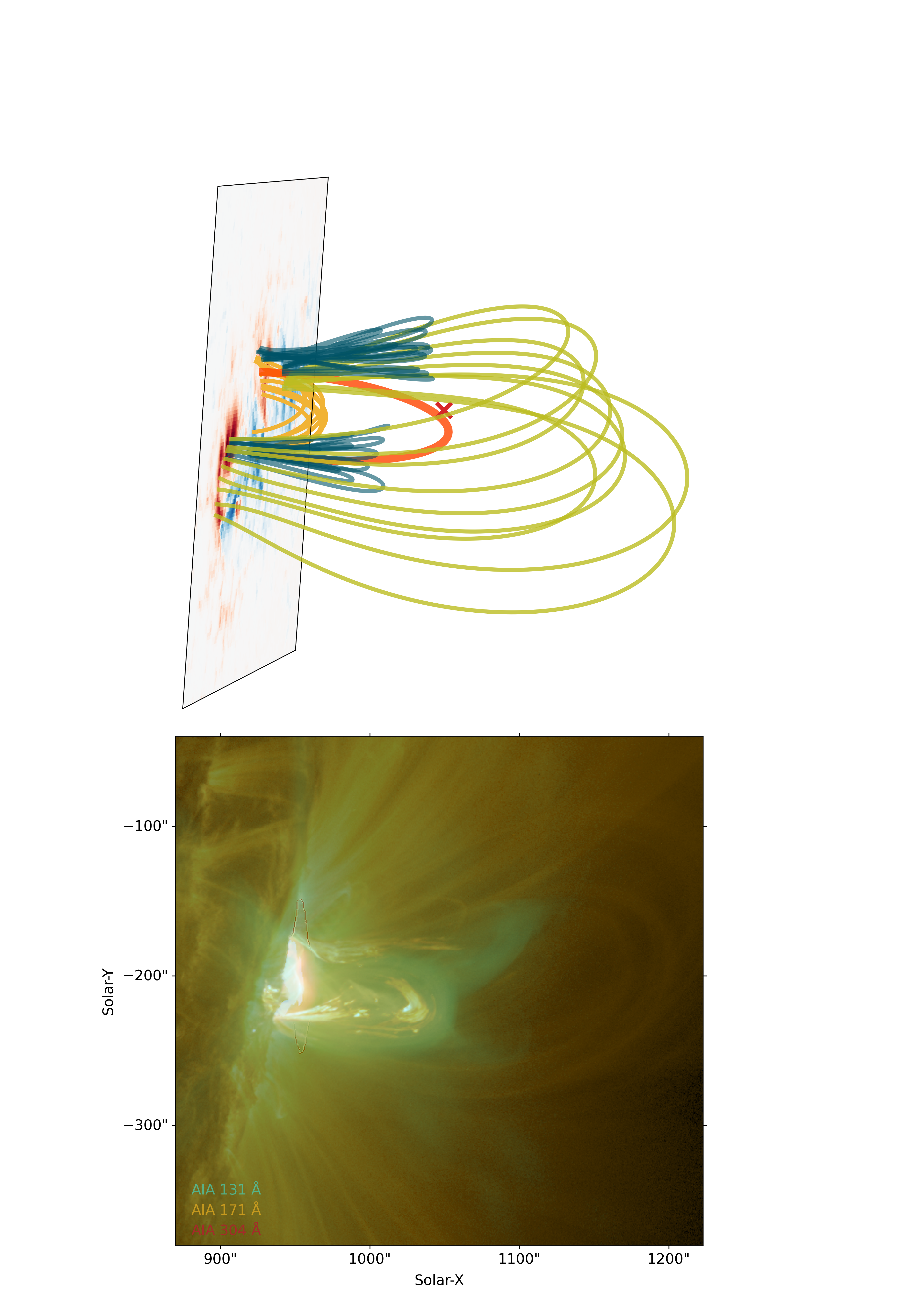}
    \caption{\small \textbf{Comparison of magnetic configuration with AIA observation.} The magnetic field lines in light orange, blue and yellow are from a potential field extrapolation (Methods; see also Fig.~\ref{fig:f2_AR}), which are rotated to a side view similar to Earth's perspective. The thick orange curve shows the 3D model of the leading front of the erupting flux rope (Methods; see also Fig~\ref{fig:f3_rec}). The two cusp-shaped outflows above the flux rope in AIA, which show Doppler red and blue shifts in EIS correspondingly, agree well with the locations of two groups of magnetic field lines in dark blue (note that the magnetic field lines are from a potential field solution, so they are different in height and morphology from the newly-reconnected outflow loops in observation).}
\end{figure}

\begin{figure}[htbp]
	\centering
	\includegraphics[width=\textwidth]{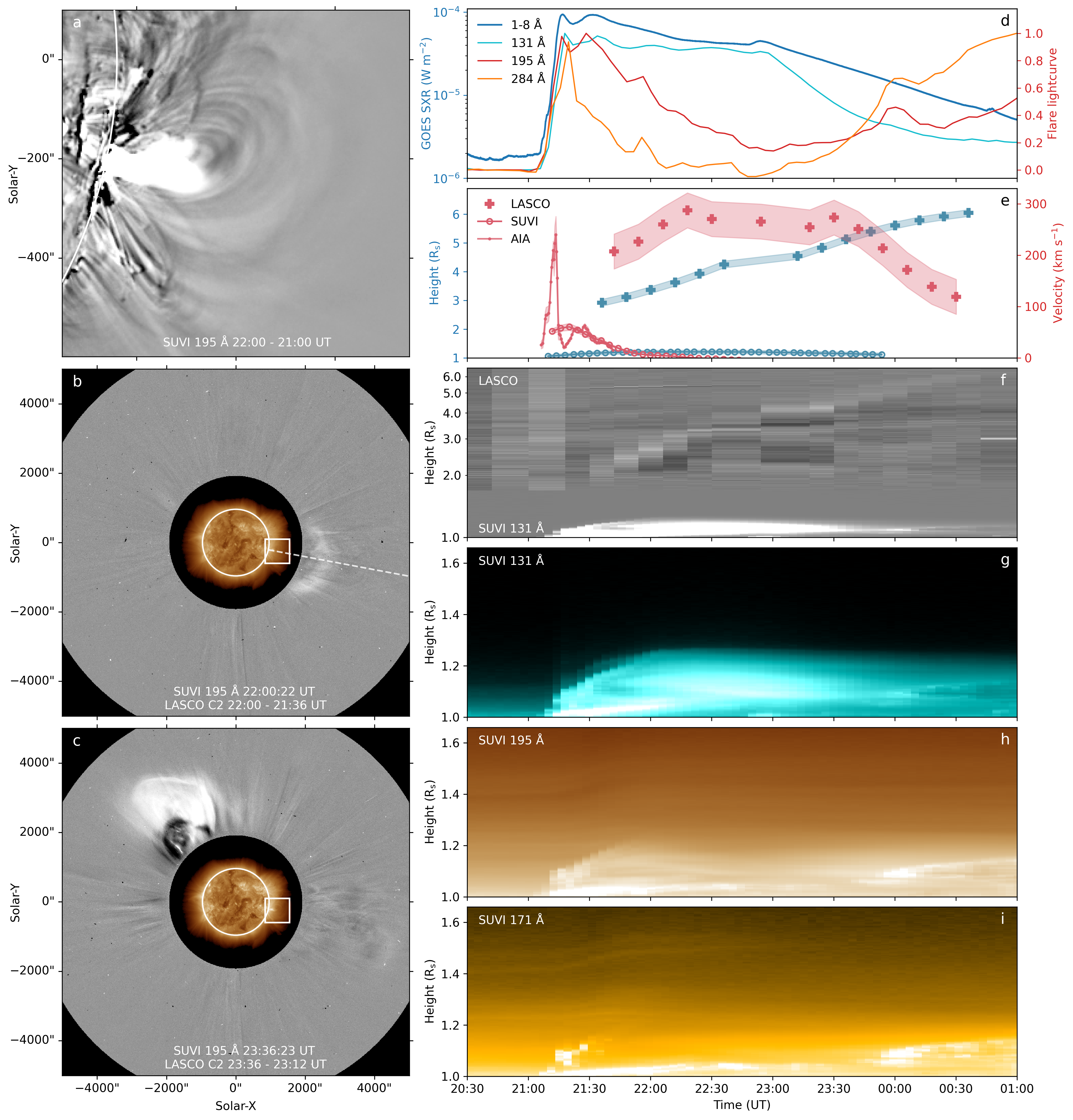}
    \caption{\small \textbf{Observations in the extended corona.} a, GOES/SUVI 195~\AA\ base difference image, showing no significant coronal dimming and almost undisturbed overlying loops after the flare. b\&c, Composite SUVI and LASCO C2 images, showing a quickly dissipating front. The white rectangle indicates the flare region, i.e., the FOV in a. The white circle indicates the solar limb. d, GOES SXR flux and normalized flare light curves (obtained within the FOV in a). e, Temporal evolution of the measured height (blue; scaled by the left y-axis) and derived velocity (red; scaled by the right y-axis). The measurements in AIA 131~\AA\ (dot symbols; see also Extended Data Fig.~2) have a much higher cadence than those in SUVI 131~\AA\ (hollow circular symbols; 4-min cadence). The velocity uncertainties are derived through error propagation of height measurement uncertainties. f--i, Height-time evolution as seen through a virtual slit above the solar limb (the dashed line in b, an extension of the dashed line in Extended Data Fig.~1). The y-axis in f is plotted in a logarithmic scale for a better visualization.}
\end{figure}


\clearpage

\begin{thebibliography}{10}
\expandafter\ifx\csname url\endcsname\relax
  \def\url#1{\texttt{#1}}\fi
\expandafter\ifx\csname urlprefix\endcsname\relax\def\urlprefix{URL }\fi
\providecommand{\bibinfo}[2]{#2}

\bibitem{Priest2002}
\bibinfo{author}{{Priest}, E.~R.} \& \bibinfo{author}{{Forbes}, T.~G.}
\newblock \bibinfo{title}{{The magnetic nature of solar flares}}.
\newblock \emph{\bibinfo{journal}{\aapr}} \textbf{\bibinfo{volume}{10}}, \bibinfo{pages}{313--377} (\bibinfo{year}{2002}).

\bibitem{Chen2011}
\bibinfo{author}{{Chen}, P.~F.}
\newblock \bibinfo{title}{{Coronal Mass Ejections: Models and Their Observational Basis}}.
\newblock \emph{\bibinfo{journal}{\lrsp}} \textbf{\bibinfo{volume}{8}}, \bibinfo{pages}{1} (\bibinfo{year}{2011}).

\bibitem{Webb2012}
\bibinfo{author}{{Webb}, D.~F.} \& \bibinfo{author}{{Howard}, T.~A.}
\newblock \bibinfo{title}{{Coronal Mass Ejections: Observations}}.
\newblock \emph{\bibinfo{journal}{\lrsp}} \textbf{\bibinfo{volume}{9}}, \bibinfo{pages}{3} (\bibinfo{year}{2012}).

\bibitem{Airapetian2020}
\bibinfo{author}{{Airapetian}, V.~S.} \emph{et~al.}
\newblock \bibinfo{title}{{Impact of space weather on climate and habitability of terrestrial-type exoplanets}}.
\newblock \emph{\bibinfo{journal}{\ija}} \textbf{\bibinfo{volume}{19}}, \bibinfo{pages}{136--194} (\bibinfo{year}{2020}).

\bibitem{lugaz2016}
\bibinfo{author}{{Lugaz}, N.} \emph{et~al.}
\newblock \bibinfo{title}{{Earth's magnetosphere and outer radiation belt under sub-Alfv{\'e}nic solar wind}}.
\newblock \emph{\bibinfo{journal}{\ncomms}} \textbf{\bibinfo{volume}{7}}, \bibinfo{pages}{13001} (\bibinfo{year}{2016}).

\bibitem{Davenport2016}
\bibinfo{author}{{Davenport}, J. R.~A.}
\newblock \bibinfo{title}{{The Kepler Catalog of Stellar Flares}}.
\newblock \emph{\bibinfo{journal}{\apj}} \textbf{\bibinfo{volume}{829}}, \bibinfo{pages}{23} (\bibinfo{year}{2016}).

\bibitem{Vida2019}
\bibinfo{author}{{Vida}, K.} \emph{et~al.}
\newblock \bibinfo{title}{{The quest for stellar coronal mass ejections in late-type stars. I. Investigating Balmer-line asymmetries of single stars in Virtual Observatory data}}.
\newblock \emph{\bibinfo{journal}{\aap}} \textbf{\bibinfo{volume}{623}}, \bibinfo{pages}{A49} (\bibinfo{year}{2019}).

\bibitem{Veronig2021}
\bibinfo{author}{{Veronig}, A.~M.} \emph{et~al.}
\newblock \bibinfo{title}{{Indications of stellar coronal mass ejections through coronal dimmings}}.
\newblock \emph{\bibinfo{journal}{\nastro}} \textbf{\bibinfo{volume}{5}}, \bibinfo{pages}{697--706} (\bibinfo{year}{2021}).

\bibitem{Odert2017}
\bibinfo{author}{{Odert}, P.}, \bibinfo{author}{{Leitzinger}, M.}, \bibinfo{author}{{Hanslmeier}, A.} \& \bibinfo{author}{{Lammer}, H.}
\newblock \bibinfo{title}{{Stellar coronal mass ejections - I. Estimating occurrence frequencies and mass-loss rates}}.
\newblock \emph{\bibinfo{journal}{\mnras}} \textbf{\bibinfo{volume}{472}}, \bibinfo{pages}{876--890} (\bibinfo{year}{2017}).

\bibitem{Green2018}
\bibinfo{author}{{Green}, L.~M.}, \bibinfo{author}{{T{\"o}r{\"o}k}, T.}, \bibinfo{author}{{Vr{\v{s}}nak}, B.}, \bibinfo{author}{{Manchester}, W.} \& \bibinfo{author}{{Veronig}, A.}
\newblock \bibinfo{title}{{The Origin, Early Evolution and Predictability of Solar Eruptions}}.
\newblock \emph{\bibinfo{journal}{\ssr}} \textbf{\bibinfo{volume}{214}}, \bibinfo{pages}{46} (\bibinfo{year}{2018}).

\bibitem{Moore2001}
\bibinfo{author}{{Moore}, R.~L.}, \bibinfo{author}{{Sterling}, A.~C.}, \bibinfo{author}{{Hudson}, H.~S.} \& \bibinfo{author}{{Lemen}, J.~R.}
\newblock \bibinfo{title}{{Onset of the Magnetic Explosion in Solar Flares and Coronal Mass Ejections}}.
\newblock \emph{\bibinfo{journal}{\apj}} \textbf{\bibinfo{volume}{552}}, \bibinfo{pages}{833--848} (\bibinfo{year}{2001}).

\bibitem{Antiochos1999}
\bibinfo{author}{{Antiochos}, S.~K.}, \bibinfo{author}{{DeVore}, C.~R.} \& \bibinfo{author}{{Klimchuk}, J.~A.}
\newblock \bibinfo{title}{{A Model for Solar Coronal Mass Ejections}}.
\newblock \emph{\bibinfo{journal}{\apj}} \textbf{\bibinfo{volume}{510}}, \bibinfo{pages}{485--493} (\bibinfo{year}{1999}).

\bibitem{Toeroek2004}
\bibinfo{author}{{T{\"o}r{\"o}k}, T.}, \bibinfo{author}{{Kliem}, B.} \& \bibinfo{author}{{Titov}, V.~S.}
\newblock \bibinfo{title}{{Ideal kink instability of a magnetic loop equilibrium}}.
\newblock \emph{\bibinfo{journal}{\aap}} \textbf{\bibinfo{volume}{413}}, \bibinfo{pages}{L27--L30} (\bibinfo{year}{2004}).

\bibitem{Gou2019}
\bibinfo{author}{{Gou}, T.}, \bibinfo{author}{{Liu}, R.}, \bibinfo{author}{{Kliem}, B.}, \bibinfo{author}{{Wang}, Y.} \& \bibinfo{author}{{Veronig}, A.~M.}
\newblock \bibinfo{title}{{The Birth of A Coronal Mass Ejection}}.
\newblock \emph{\bibinfo{journal}{\sciadv}} \textbf{\bibinfo{volume}{5}}, \bibinfo{pages}{7004} (\bibinfo{year}{2019}).

\bibitem{Liu2020}
\bibinfo{author}{{Liu}, R.}
\newblock \bibinfo{title}{{Magnetic flux ropes in the solar corona: structure and evolution toward eruption}}.
\newblock \emph{\bibinfo{journal}{\raa}} \textbf{\bibinfo{volume}{20}}, \bibinfo{pages}{165} (\bibinfo{year}{2020}).

\bibitem{Lin2000}
\bibinfo{author}{{Lin}, J.} \& \bibinfo{author}{{Forbes}, T.~G.}
\newblock \bibinfo{title}{{Effects of reconnection on the coronal mass ejection process}}.
\newblock \emph{\bibinfo{journal}{\jgr}} \textbf{\bibinfo{volume}{105}}, \bibinfo{pages}{2375--2392} (\bibinfo{year}{2000}).

\bibitem{Gou2020}
\bibinfo{author}{{Gou}, T.} \emph{et~al.}
\newblock \bibinfo{title}{{Solar Flare-CME Coupling throughout Two Acceleration Phases of a Fast CME}}.
\newblock \emph{\bibinfo{journal}{\apjl}} \textbf{\bibinfo{volume}{897}}, \bibinfo{pages}{L36} (\bibinfo{year}{2020}).

\bibitem{Bateman1978}
\bibinfo{author}{{Bateman}, G.}
\newblock \emph{\bibinfo{title}{{MHD instabilities}}} (\bibinfo{year}{1978}).

\bibitem{Kliem2006}
\bibinfo{author}{{Kliem}, B.} \& \bibinfo{author}{{T{\"o}r{\"o}k}, T.}
\newblock \bibinfo{title}{{Torus Instability}}.
\newblock \emph{\bibinfo{journal}{\prl}} \textbf{\bibinfo{volume}{96}}, \bibinfo{pages}{255002} (\bibinfo{year}{2006}).

\bibitem{Falconer2002}
\bibinfo{author}{{Falconer}, D.~A.}, \bibinfo{author}{{Moore}, R.~L.} \& \bibinfo{author}{{Gary}, G.~A.}
\newblock \bibinfo{title}{{Correlation of the Coronal Mass Ejection Productivity of Solar Active Regions with Measures of Their Global Nonpotentiality from Vector Magnetograms: Baseline Results}}.
\newblock \emph{\bibinfo{journal}{\apj}} \textbf{\bibinfo{volume}{569}}, \bibinfo{pages}{1016--1025} (\bibinfo{year}{2002}).

\bibitem{Wang2007}
\bibinfo{author}{{Wang}, Y.} \& \bibinfo{author}{{Zhang}, J.}
\newblock \bibinfo{title}{{A Comparative Study between Eruptive X-Class Flares Associated with Coronal Mass Ejections and Confined X-Class Flares}}.
\newblock \emph{\bibinfo{journal}{\apj}} \textbf{\bibinfo{volume}{665}}, \bibinfo{pages}{1428--1438} (\bibinfo{year}{2007}).

\bibitem{Sun2015}
\bibinfo{author}{{Sun}, X.} \emph{et~al.}
\newblock \bibinfo{title}{{Why Is the Great Solar Active Region 12192 Flare-rich but CME-poor?}}
\newblock \emph{\bibinfo{journal}{\apjl}} \textbf{\bibinfo{volume}{804}}, \bibinfo{pages}{L28} (\bibinfo{year}{2015}).

\bibitem{Torok2005}
\bibinfo{author}{{T{\"o}r{\"o}k}, T.} \& \bibinfo{author}{{Kliem}, B.}
\newblock \bibinfo{title}{{Confined and Ejective Eruptions of Kink-unstable Flux Ropes}}.
\newblock \emph{\bibinfo{journal}{\apjl}} \textbf{\bibinfo{volume}{630}}, \bibinfo{pages}{L97--L100} (\bibinfo{year}{2005}).

\bibitem{Fan2007}
\bibinfo{author}{{Fan}, Y.} \& \bibinfo{author}{{Gibson}, S.~E.}
\newblock \bibinfo{title}{{Onset of Coronal Mass Ejections Due to Loss of Confinement of Coronal Flux Ropes}}.
\newblock \emph{\bibinfo{journal}{\apj}} \textbf{\bibinfo{volume}{668}}, \bibinfo{pages}{1232--1245} (\bibinfo{year}{2007}).

\bibitem{Ji2003}
\bibinfo{author}{{Ji}, H.}, \bibinfo{author}{{Wang}, H.}, \bibinfo{author}{{Schmahl}, E.~J.}, \bibinfo{author}{{Moon}, Y.~J.} \& \bibinfo{author}{{Jiang}, Y.}
\newblock \bibinfo{title}{{Observations of the Failed Eruption of a Filament}}.
\newblock \emph{\bibinfo{journal}{\apjl}} \textbf{\bibinfo{volume}{595}}, \bibinfo{pages}{L135--L138} (\bibinfo{year}{2003}).

\bibitem{Liu2009}
\bibinfo{author}{{Liu}, Y.} \emph{et~al.}
\newblock \bibinfo{title}{{New Observation of Failed Filament Eruptions: The Influence of Asymmetric Coronal Background Fields on Solar Eruptions}}.
\newblock \emph{\bibinfo{journal}{\apjl}} \textbf{\bibinfo{volume}{696}}, \bibinfo{pages}{L70--L73} (\bibinfo{year}{2009}).

\bibitem{Chen2023}
\bibinfo{author}{{Chen}, Y.}, \bibinfo{author}{{Cheng}, X.}, \bibinfo{author}{{Chen}, J.}, \bibinfo{author}{{Dai}, Y.} \& \bibinfo{author}{{Ding}, M.}
\newblock \bibinfo{title}{{Observations of a Failed Solar Filament Eruption Involving External Reconnection}}.
\newblock \emph{\bibinfo{journal}{\apj}} \textbf{\bibinfo{volume}{959}}, \bibinfo{pages}{67} (\bibinfo{year}{2023}).

\bibitem{Karpen2024}
\bibinfo{author}{{Karpen}, J.~T.}, \bibinfo{author}{{Kumar}, P.}, \bibinfo{author}{{Wyper}, P.~F.}, \bibinfo{author}{{DeVore}, C.~R.} \& \bibinfo{author}{{Antiochos}, S.~K.}
\newblock \bibinfo{title}{{Solar Eruptions in Nested Magnetic Flux Systems}}.
\newblock \emph{\bibinfo{journal}{\apj}} \textbf{\bibinfo{volume}{966}}, \bibinfo{pages}{27} (\bibinfo{year}{2024}).

\bibitem{gibson2006}
\bibinfo{author}{{Gibson}, S.~E.} \& \bibinfo{author}{{Fan}, Y.}
\newblock \bibinfo{title}{{Coronal prominence structure and dynamics: A magnetic flux rope interpretation}}.
\newblock \emph{\bibinfo{journal}{\jgr}} \textbf{\bibinfo{volume}{111}}, \bibinfo{pages}{A12103} (\bibinfo{year}{2006}).

\bibitem{Zhang2012}
\bibinfo{author}{{Zhang}, J.}, \bibinfo{author}{{Cheng}, X.} \& \bibinfo{author}{{Ding}, M.-D.}
\newblock \bibinfo{title}{{Observation of an evolving magnetic flux rope before and during a solar eruption}}.
\newblock \emph{\bibinfo{journal}{\ncomms}} \textbf{\bibinfo{volume}{3}}, \bibinfo{pages}{747} (\bibinfo{year}{2012}).

\bibitem{Lynch2008}
\bibinfo{author}{{Lynch}, B.~J.}, \bibinfo{author}{{Antiochos}, S.~K.}, \bibinfo{author}{{DeVore}, C.~R.}, \bibinfo{author}{{Luhmann}, J.~G.} \& \bibinfo{author}{{Zurbuchen}, T.~H.}
\newblock \bibinfo{title}{{Topological Evolution of a Fast Magnetic Breakout CME in Three Dimensions}}.
\newblock \emph{\bibinfo{journal}{\apj}} \textbf{\bibinfo{volume}{683}}, \bibinfo{pages}{1192--1206} (\bibinfo{year}{2008}).

\bibitem{Shibata2011}
\bibinfo{author}{{Shibata}, K.} \& \bibinfo{author}{{Magara}, T.}
\newblock \bibinfo{title}{{Solar Flares: Magnetohydrodynamic Processes}}.
\newblock \emph{\bibinfo{journal}{\lrsp}} \textbf{\bibinfo{volume}{8}}, \bibinfo{pages}{6} (\bibinfo{year}{2011}).

\bibitem{Karpen2012}
\bibinfo{author}{{Karpen}, J.~T.}, \bibinfo{author}{{Antiochos}, S.~K.} \& \bibinfo{author}{{DeVore}, C.~R.}
\newblock \bibinfo{title}{{The Mechanisms for the Onset and Explosive Eruption of Coronal Mass Ejections and Eruptive Flares}}.
\newblock \emph{\bibinfo{journal}{\apj}} \textbf{\bibinfo{volume}{760}}, \bibinfo{pages}{81} (\bibinfo{year}{2012}).

\bibitem{Petschek1964}
\bibinfo{author}{{Petschek}, H.~E.}
\newblock \bibinfo{title}{{Magnetic Field Annihilation}}.
\newblock In \bibinfo{editor}{{Hess}, W.~N.} (ed.) \emph{\bibinfo{booktitle}{NASA Special Publication}}, vol.~\bibinfo{volume}{50}, \bibinfo{pages}{425} (\bibinfo{year}{1964}).

\bibitem{Lin2005}
\bibinfo{author}{{Lin}, J.} \emph{et~al.}
\newblock \bibinfo{title}{{Direct Observations of the Magnetic Reconnection Site of an Eruption on 2003 November 18}}.
\newblock \emph{\bibinfo{journal}{\apj}} \textbf{\bibinfo{volume}{622}}, \bibinfo{pages}{1251--1264} (\bibinfo{year}{2005}).

\bibitem{Gou2017}
\bibinfo{author}{{Gou}, T.}, \bibinfo{author}{{Veronig}, A.~M.}, \bibinfo{author}{{Dickson}, E.~C.}, \bibinfo{author}{{Hernandez-Perez}, A.} \& \bibinfo{author}{{Liu}, R.}
\newblock \bibinfo{title}{{Direct Observation of Two-step Magnetic Reconnection in a Solar Flare}}.
\newblock \emph{\bibinfo{journal}{\apjl}} \textbf{\bibinfo{volume}{845}}, \bibinfo{pages}{L1} (\bibinfo{year}{2017}).

\bibitem{Vrsnak2008}
\bibinfo{author}{{Vr{\v{s}}nak}, B.}
\newblock \bibinfo{title}{{Processes and mechanisms governing the initiation and propagation of CMEs}}.
\newblock \emph{\bibinfo{journal}{\ag}} \textbf{\bibinfo{volume}{26}}, \bibinfo{pages}{3089--3101} (\bibinfo{year}{2008}).

\bibitem{Zhang2001}
\bibinfo{author}{{Zhang}, J.}, \bibinfo{author}{{Dere}, K.~P.}, \bibinfo{author}{{Howard}, R.~A.}, \bibinfo{author}{{Kundu}, M.~R.} \& \bibinfo{author}{{White}, S.~M.}
\newblock \bibinfo{title}{{On the Temporal Relationship between Coronal Mass Ejections and Flares}}.
\newblock \emph{\bibinfo{journal}{\apj}} \textbf{\bibinfo{volume}{559}}, \bibinfo{pages}{452--462} (\bibinfo{year}{2001}).

\bibitem{Temmer2010}
\bibinfo{author}{{Temmer}, M.}, \bibinfo{author}{{Veronig}, A.~M.}, \bibinfo{author}{{Kontar}, E.~P.}, \bibinfo{author}{{Krucker}, S.} \& \bibinfo{author}{{Vr{\v{s}}nak}, B.}
\newblock \bibinfo{title}{{Combined STEREO/RHESSI Study of Coronal Mass Ejection Acceleration and Particle Acceleration in Solar Flares}}.
\newblock \emph{\bibinfo{journal}{\apj}} \textbf{\bibinfo{volume}{712}}, \bibinfo{pages}{1410--1420} (\bibinfo{year}{2010}).

\bibitem{Aulanier2019}
\bibinfo{author}{{Aulanier}, G.} \& \bibinfo{author}{{Dud{\'\i}k}, J.}
\newblock \bibinfo{title}{{Drifting of the line-tied footpoints of CME flux-ropes}}.
\newblock \emph{\bibinfo{journal}{\aap}} \textbf{\bibinfo{volume}{621}}, \bibinfo{pages}{A72} (\bibinfo{year}{2019}).

\bibitem{Jiang2023}
\bibinfo{author}{{Jiang}, C.} \emph{et~al.}
\newblock \bibinfo{title}{{A model of failed solar eruption initiated and destructed by magnetic reconnection}}.
\newblock \emph{\bibinfo{journal}{\mnras}} \textbf{\bibinfo{volume}{525}}, \bibinfo{pages}{5857--5867} (\bibinfo{year}{2023}).

\bibitem{Gou2023}
\bibinfo{author}{{Gou}, T.} \emph{et~al.}
\newblock \bibinfo{title}{{Complete replacement of magnetic flux in a flux rope during a coronal mass ejection}}.
\newblock \emph{\bibinfo{journal}{\nastro}} \textbf{\bibinfo{volume}{7}}, \bibinfo{pages}{815--824} (\bibinfo{year}{2023}).

\bibitem{Ruffenach2012}
\bibinfo{author}{{Ruffenach}, A.} \emph{et~al.}
\newblock \bibinfo{title}{{Multispacecraft observation of magnetic cloud erosion by magnetic reconnection during propagation}}.
\newblock \emph{\bibinfo{journal}{\jgr}} \textbf{\bibinfo{volume}{117}}, \bibinfo{pages}{A09101} (\bibinfo{year}{2012}).

\bibitem{Luo2022}
\bibinfo{author}{{Luo}, R.} \& \bibinfo{author}{{Liu}, R.}
\newblock \bibinfo{title}{{Where and How Does a Decay-index Profile Become Saddle-like?}}
\newblock \emph{\bibinfo{journal}{\apj}} \textbf{\bibinfo{volume}{929}}, \bibinfo{pages}{2} (\bibinfo{year}{2022}).

\bibitem{DeVore2008}
\bibinfo{author}{{DeVore}, C.~R.} \& \bibinfo{author}{{Antiochos}, S.~K.}
\newblock \bibinfo{title}{{Homologous Confined Filament Eruptions via Magnetic Breakout}}.
\newblock \emph{\bibinfo{journal}{\apj}} \textbf{\bibinfo{volume}{680}}, \bibinfo{pages}{740--756} (\bibinfo{year}{2008}).

\bibitem{Demoulin1996}
\bibinfo{author}{{Demoulin}, P.}, \bibinfo{author}{{Henoux}, J.~C.}, \bibinfo{author}{{Priest}, E.~R.} \& \bibinfo{author}{{Mandrini}, C.~H.}
\newblock \bibinfo{title}{{Quasi-Separatrix layers in solar flares. I. Method.}}
\newblock \emph{\bibinfo{journal}{\aap}} \textbf{\bibinfo{volume}{308}}, \bibinfo{pages}{643--655} (\bibinfo{year}{1996}).

\bibitem{Yashiro2006}
\bibinfo{author}{{Yashiro}, S.}, \bibinfo{author}{{Akiyama}, S.}, \bibinfo{author}{{Gopalswamy}, N.} \& \bibinfo{author}{{Howard}, R.~A.}
\newblock \bibinfo{title}{{Different Power-Law Indices in the Frequency Distributions of Flares with and without Coronal Mass Ejections}}.
\newblock \emph{\bibinfo{journal}{\apjl}} \textbf{\bibinfo{volume}{650}}, \bibinfo{pages}{L143--L146} (\bibinfo{year}{2006}).

\bibitem{Li2020}
\bibinfo{author}{{Li}, T.} \emph{et~al.}
\newblock \bibinfo{title}{{Magnetic Flux of Active Regions Determining the Eruptive Character of Large Solar Flares}}.
\newblock \emph{\bibinfo{journal}{\apj}} \textbf{\bibinfo{volume}{900}}, \bibinfo{pages}{128} (\bibinfo{year}{2020}).

\bibitem{Donati2009}
\bibinfo{author}{{Donati}, J.~F.} \& \bibinfo{author}{{Landstreet}, J.~D.}
\newblock \bibinfo{title}{{Magnetic Fields of Nondegenerate Stars}}.
\newblock \emph{\bibinfo{journal}{\araa}} \textbf{\bibinfo{volume}{47}}, \bibinfo{pages}{333--370} (\bibinfo{year}{2009}).

\bibitem{AlvaradoGomez2018}
\bibinfo{author}{{Alvarado-G{\'o}mez}, J.~D.}, \bibinfo{author}{{Drake}, J.~J.}, \bibinfo{author}{{Cohen}, O.}, \bibinfo{author}{{Moschou}, S.~P.} \& \bibinfo{author}{{Garraffo}, C.}
\newblock \bibinfo{title}{{Suppression of Coronal Mass Ejections in Active Stars by an Overlying Large-scale Magnetic Field: A Numerical Study}}.
\newblock \emph{\bibinfo{journal}{\apj}} \textbf{\bibinfo{volume}{862}}, \bibinfo{pages}{93} (\bibinfo{year}{2018}).

\bibitem{Lemen2012}
\bibinfo{author}{{Lemen}, J.~R.} \emph{et~al.}
\newblock \bibinfo{title}{{The Atmospheric Imaging Assembly (AIA) on the Solar Dynamics Observatory (SDO)}}.
\newblock \emph{\bibinfo{journal}{\solphys}} \textbf{\bibinfo{volume}{275}}, \bibinfo{pages}{17--40} (\bibinfo{year}{2012}).

\bibitem{Pesnell2012}
\bibinfo{author}{{Pesnell}, W.~D.}, \bibinfo{author}{{Thompson}, B.~J.} \& \bibinfo{author}{{Chamberlin}, P.~C.}
\newblock \bibinfo{title}{{The Solar Dynamics Observatory (SDO)}}.
\newblock \emph{\bibinfo{journal}{\solphys}} \textbf{\bibinfo{volume}{275}}, \bibinfo{pages}{3--15} (\bibinfo{year}{2012}).

\bibitem{Cheung2015}
\bibinfo{author}{{Cheung}, M. C.~M.} \emph{et~al.}
\newblock \bibinfo{title}{{Thermal Diagnostics with the Atmospheric Imaging Assembly on board the Solar Dynamics Observatory: A Validated Method for Differential Emission Measure Inversions}}.
\newblock \emph{\bibinfo{journal}{\apj}} \textbf{\bibinfo{volume}{807}}, \bibinfo{pages}{143} (\bibinfo{year}{2015}).

\bibitem{Golub2007}
\bibinfo{author}{{Golub}, L.} \emph{et~al.}
\newblock \bibinfo{title}{{The X-Ray Telescope (XRT) for the Hinode Mission}}.
\newblock \emph{\bibinfo{journal}{\solphys}} \textbf{\bibinfo{volume}{243}}, \bibinfo{pages}{63--86} (\bibinfo{year}{2007}).

\bibitem{Mueller2020}
\bibinfo{author}{{M{\"u}ller}, D.} \emph{et~al.}
\newblock \bibinfo{title}{{The Solar Orbiter mission. Science overview}}.
\newblock \emph{\bibinfo{journal}{\aap}} \textbf{\bibinfo{volume}{642}}, \bibinfo{pages}{A1} (\bibinfo{year}{2020}).

\bibitem{Rochus2020}
\bibinfo{author}{{Rochus}, P.} \emph{et~al.}
\newblock \bibinfo{title}{{The Solar Orbiter EUI instrument: The Extreme Ultraviolet Imager}}.
\newblock \emph{\bibinfo{journal}{\aap}} \textbf{\bibinfo{volume}{642}}, \bibinfo{pages}{A8} (\bibinfo{year}{2020}).

\bibitem{Veronig2025}
\bibinfo{author}{{Veronig}, A.~M.} \emph{et~al.}
\newblock \bibinfo{title}{{Coronal dimmings and what they tell us about solar and stellar coronal mass ejections}}.
\newblock \emph{\bibinfo{journal}{\lrsp}} \textbf{\bibinfo{volume}{22}}, \bibinfo{pages}{2} (\bibinfo{year}{2025}).

\bibitem{Brueckner1995}
\bibinfo{author}{{Brueckner}, G.~E.} \emph{et~al.}
\newblock \bibinfo{title}{{The Large Angle Spectroscopic Coronagraph (LASCO)}}.
\newblock \emph{\bibinfo{journal}{\solphys}} \textbf{\bibinfo{volume}{162}}, \bibinfo{pages}{357--402} (\bibinfo{year}{1995}).

\bibitem{Cargill2004}
\bibinfo{author}{{Cargill}, P.~J.}
\newblock \bibinfo{title}{{On the Aerodynamic Drag Force Acting on Interplanetary Coronal Mass Ejections}}.
\newblock \emph{\bibinfo{journal}{\solphys}} \textbf{\bibinfo{volume}{221}}, \bibinfo{pages}{135--149} (\bibinfo{year}{2004}).

\bibitem{Vrsnak2007}
\bibinfo{author}{{Vr{\v{s}}nak}, B.} \& \bibinfo{author}{{{\v{Z}}ic}, T.}
\newblock \bibinfo{title}{{Transit times of interplanetary coronal mass ejections and the solar wind speed}}.
\newblock \emph{\bibinfo{journal}{\aap}} \textbf{\bibinfo{volume}{472}}, \bibinfo{pages}{937--943} (\bibinfo{year}{2007}).

\bibitem{Warmuth2015}
\bibinfo{author}{{Warmuth}, A.}
\newblock \bibinfo{title}{{Large-scale Globally Propagating Coronal Waves}}.
\newblock \emph{\bibinfo{journal}{\lrsp}} \textbf{\bibinfo{volume}{12}}, \bibinfo{pages}{3} (\bibinfo{year}{2015}).

\bibitem{Howard2016}
\bibinfo{author}{{Howard}, T.~A.} \& \bibinfo{author}{{Pizzo}, V.~J.}
\newblock \bibinfo{title}{{Challenging Some Contemporary Views of Coronal Mass Ejections. I. The Case for Blast Waves}}.
\newblock \emph{\bibinfo{journal}{\apj}} \textbf{\bibinfo{volume}{824}}, \bibinfo{pages}{92} (\bibinfo{year}{2016}).

\bibitem{Morosan2023}
\bibinfo{author}{{Morosan}, D.~E.}, \bibinfo{author}{{Pomoell}, J.}, \bibinfo{author}{{Kumari}, A.}, \bibinfo{author}{{Kilpua}, E.~K.~J.} \& \bibinfo{author}{{Vainio}, R.}
\newblock \bibinfo{title}{{A type II solar radio burst without a coronal mass ejection}}.
\newblock \emph{\bibinfo{journal}{\aap}} \textbf{\bibinfo{volume}{675}}, \bibinfo{pages}{A98} (\bibinfo{year}{2023}).

\bibitem{Culhane2007}
\bibinfo{author}{{Culhane}, J.~L.} \emph{et~al.}
\newblock \bibinfo{title}{{The EUV Imaging Spectrometer for Hinode}}.
\newblock \emph{\bibinfo{journal}{\solphys}} \textbf{\bibinfo{volume}{243}}, \bibinfo{pages}{19--61} (\bibinfo{year}{2007}).

\bibitem{DePontieu2014}
\bibinfo{author}{{De Pontieu}, B.} \emph{et~al.}
\newblock \bibinfo{title}{{The Interface Region Imaging Spectrograph (IRIS)}}.
\newblock \emph{\bibinfo{journal}{\solphys}} \textbf{\bibinfo{volume}{289}}, \bibinfo{pages}{2733--2779} (\bibinfo{year}{2014}).

\bibitem{Krucker2020}
\bibinfo{author}{{Krucker}, S.} \emph{et~al.}
\newblock \bibinfo{title}{{The Spectrometer/Telescope for Imaging X-rays (STIX)}}.
\newblock \emph{\bibinfo{journal}{\aap}} \textbf{\bibinfo{volume}{642}}, \bibinfo{pages}{A15} (\bibinfo{year}{2020}).

\bibitem{Massa2023}
\bibinfo{author}{{Massa}, P.} \emph{et~al.}
\newblock \bibinfo{title}{{The STIX Imaging Concept}}.
\newblock \emph{\bibinfo{journal}{\solphys}} \textbf{\bibinfo{volume}{298}}, \bibinfo{pages}{114} (\bibinfo{year}{2023}).

\bibitem{Gary2018}
\bibinfo{author}{{Gary}, D.~E.} \emph{et~al.}
\newblock \bibinfo{title}{{Microwave and Hard X-Ray Observations of the 2017 September 10 Solar Limb Flare}}.
\newblock \emph{\bibinfo{journal}{\apj}} \textbf{\bibinfo{volume}{863}}, \bibinfo{pages}{83} (\bibinfo{year}{2018}).

\bibitem{Scherrer2012}
\bibinfo{author}{{Scherrer}, P.~H.} \emph{et~al.}
\newblock \bibinfo{title}{{The Helioseismic and Magnetic Imager (HMI) Investigation for the Solar Dynamics Observatory (SDO)}}.
\newblock \emph{\bibinfo{journal}{\solphys}} \textbf{\bibinfo{volume}{275}}, \bibinfo{pages}{207--227} (\bibinfo{year}{2012}).

\bibitem{Solanki2020}
\bibinfo{author}{{Solanki}, S.~K.} \emph{et~al.}
\newblock \bibinfo{title}{{The Polarimetric and Helioseismic Imager on Solar Orbiter}}.
\newblock \emph{\bibinfo{journal}{\aap}} \textbf{\bibinfo{volume}{642}}, \bibinfo{pages}{A11} (\bibinfo{year}{2020}).

\bibitem{Alissandrakis1981}
\bibinfo{author}{{Alissandrakis}, C.~E.}
\newblock \bibinfo{title}{{On the computation of constant alpha force-free magnetic field}}.
\newblock \emph{\bibinfo{journal}{\aap}} \textbf{\bibinfo{volume}{100}}, \bibinfo{pages}{197--200} (\bibinfo{year}{1981}).

\bibitem{Wuelser2004}
\bibinfo{author}{{Wuelser}, J.-P.} \emph{et~al.}
\newblock \bibinfo{title}{{EUVI: the STEREO-SECCHI extreme ultraviolet imager}}.
\newblock In \bibinfo{editor}{{Fineschi}, S.} \& \bibinfo{editor}{{Gummin}, M.~A.} (eds.) \emph{\bibinfo{booktitle}{Telescopes and Instrumentation for Solar Astrophysics}}, vol. \bibinfo{volume}{5171} of \emph{\bibinfo{series}{Society of Photo-Optical Instrumentation Engineers (SPIE) Conference Series}}, \bibinfo{pages}{111--122} (\bibinfo{year}{2004}).

\end{thebibliography}
\end{document}